\documentclass[12pt,letterpaper]{article}

\usepackage{hyperref}
\usepackage{appendix}

\usepackage{amsmath, amssymb, amsthm}
\usepackage{graphicx,color}
\usepackage[left=1.5in, right=1in, top=1in, bottom=1in, includefoot, headheight=13.6pt]{geometry}
\usepackage[square, comma, numbers, sort&compress]{natbib}
\usepackage[T1]{fontenc}

\usepackage{sectsty}
\chapterfont{\large\sc\centering}
\chaptertitlefont{\centering}
\subsubsectionfont{\centering}

\numberwithin{equation}{section}

\numberwithin{equation}{section}

\def\be{\begin{equation}}
\def\ee{\end{equation}}
\def\ba{\begin{eqnarray}}
\def\ea{\end{eqnarray}}
\def\la{\label}

\title{\Huge\textbf{Bimetric Models of Gravity and Cosmology in the Early Universe}}
\author{Jos\'e Tom\'as G\'alvez Ghersi\\ Supervised by Dr. Carlo R. Contaldi}
\date{September 2013}
\begin{document}
\maketitle
\vspace{2cm}
\begin{center}
\rule{8.8cm}{0.3mm}\\
\begin{center}
\emph{\small{Submitted in partial fulfilment of the requirements\\
for the degree of Master of Science\\
of Imperial College London}}\\
\end{center}
\rule{8.8cm}{0.3mm}
\end{center}
\thispagestyle{empty}
\newpage
\null\vfill
\begin{abstract}
In the last decades, we have witnessed an astonishing progress in our understanding of the Early Universe and its evolution into the actual configuration we observe. General Relativity has been a key ingredient for this achievement, and it certainly provides the correct way to couple matter fields to gravity. However, in other fundamental theories (string theory, for example), we can find an immense number of degrees of freedom, and some of them are specially relevant in order to analyze cosmic evolution. Then a question arises: Is this the only way to couple these modes to gravity and matter?\\
This dissertation takes the opportunity to explore \emph{Bimetric Models of Gravity} as theories of modified gravity motivated by cosmological purposes. Such as the solutions of the issues presented on standard Big Bang Cosmology, the Horizon and the flatness problem on the matter sector. The starting point for this ideas takes into account the possibility of modified dynamics for the matter fields, and how this modifications translate into new geometrical features.\\   
In this dissertation, we explore models based on the idea that there are two metrics in spacetime: One describes the standard gravity, and the other provides a geometry in which matter fields propagate.  In order to do that, we provide the essentials of Finsler geometry and the rules to induce a metric for the propagation of matter. Such a description will cover some of the most critical features related to the field necessary to do the induction, these will arise in an attempt to build an action for this field. And finally, we provide an example to study the homogeneous limit of background {\it FLRW} equations for the cosmological model and the role of Lorentz symmetry breaking to provide a \textit{graceful exit}.\\ 
\end{abstract}
\vfill
\thispagestyle{empty}
\newpage
\tableofcontents
\thispagestyle{empty}
\newpage
\section{Introduction}
In standard General Relativity (GR) we find a coherent way to couple gravity to matter fields. However, there are not convincing arguments to think that the equations of motion for gravity do not present couplings to other fields. Many contemporary approaches predict an immense number of degrees of freedom which might be specially relevant at high energy scales, playing a major role in the early stages of the universe. Of course, the effect of such additional degrees of freedom should be supressed at the current scales (of energy and length) in which GR has been tested successfully.\\
In the last 30 years, the literature has presented a plethora of alternatives for modified gravity \cite{Clifton:2011jh}. Most of them are conceived as deviations from GR, motivated by certain physical phenomena; or in the contrary, as attempts to rule out some results that became essential in our current understanding. GR has shedded many lights on our knowledge of the actual structure of the Universe and its evolution. And it is natural to think about cosmological issues as motivations to modify gravity.\par In this essay, we explore \textit{Bimetric Gravity}, highlighting the features we can use to solve the pathologies found in the Early Universe Cosmology. As the name suggests, this model is based on the idea that there are two metrics in spacetime: One is used to describe gravitational vacuum, and the other provides a geometry in which matter fields propagate. The latter is a metric induced from the gravitational metric by introducing additional degrees of freedom coupled to gravity. According to \cite{Clayton:1998hv}, \cite{Magueijo:2007gf} this model exhibits a variable speed of light. In the last few years, intense discussions around \cite{Clayton:1998hv} and \cite{Albrecht:1998ir} proof that it is worthwhile to consider these propsals as alternatives to solve the issues of Big Bang Cosmology.
\par To do so, first, we must describe an inducted geometry in which matter fields interact. And for that purpose, the essentials of Finsler geometries will satisfy naturally our need to generate a matter geometry from \emph{superluminal} dispersion relations. Our approach for these modifications is conservative, following the results of J. Bekenstein in \cite{Bekenstein:1992pj}: the metric emerges from a modified dispersion relation that is still quadratic, and is \textit{parametrized} by external degrees of freedom.
\newpage
In advance, we should be aware that the notions of Finsler Geometry exhibit a general procedure: these motivate many other candidates for the matter dynamics since it is always possible to modify the dispersion relations at our free will. And in consequence, some assumptions are necessary to fit in the objectives of this project. Considering causal propagation of all degrees of freedom, and requiring a natural setup to write the equations of motion. An adequate set of constraints will clarify the geometrical features of this model. 
\par Then we propose an action, which includes terms for extra fields used to induce the matter geometry. If we want to be consistent with the idea of an \textit{affordable} superluminal propagation for all fields in the matter sector, the assignment of a functional for these extra degrees of freedom is not a trivial task. By considering the case of a scalar field, we explore superficially the consequences of introducing a canonical Klein-Gordon action just by analyzing its equations of motion and its prospective matter couplings.  This Klein-Gordon action is a natural limit for a generic k-essence field. A preliminar analysis will lead us to briefly discuss the arising of Chameleon screening effect from the laws of motion.\\ In addition to that, we explore the possibility of having a \textit{Cuscuton} playing the role of an inducing field. The solutions and specific properties of this entity provide fertile soil for physical interpretations. We discuss the dynamics and other peculiarities of this proposal.
\par Knowing the constraints, the mechanisms of induction and the motion equations for the fields involved, we are able to explore a cosmological scenario proposed in this setup. In order to proceed, we write the Raychaudhuri equation for a congruence of timelike curves propagating in the matter geometry. We find that it is not compulsory to consider violations of the strong energy condition in the matter sector to achieve an expanding cosmology. Nevertheless, the result certainly confirms that matter contributions will become subdominant and irrelevant for structure formation.\\  Knowing this, we translate these results into the simplest modification of an FLRW (Friedmann-Lemaitre-Robertson-Walker) model, finding an inflationary phase irrespective of the species in the matter sector. Leading us to a direct solution of the Horizon and flatness problems. This is followed by a short discussion about the consequences of the remnants of an inexact Lorentz symmetry for particle physics.
\newpage
This provides a natural solution to the Horizon and homogeneity problems of standard Big Bang Cosmology. Therefore, we are obliged to test a graceful exit scenario for this proposal. Specially since this case is significantly different from the standard inflationary paradigm \cite{Guth:1980zm}, which is nicely designed to fit well with current observations.\\
Disformal relations between the gravitational and the induced metric imply Lorentz symmetry breaking when we refer to a Minkowski background, and there is no general consensus about the way in which one must deal with these pathologies. In this project, we make an explicit approach to this issue and its implications by pointing out that small Lorentz violations can lead to unacceptable large effects. We will mention recent approaches to face these objections.
\par The plan of the dissertation is as follows. In Section \ref{secttwo} we review the method to build the induced metric and provide a general description of the geometry for the matter sector. In this way, each relevant modified dispersion relation can be (possibly) linked with an induced metric. This geometry presents a widened lightcone due to a dynamic speed of light. In Section \ref{sectthree} we set up an action and the corresponding motion equations for the fields inducing this geometry. In here, we intend to connect the dynamics of the gravitational sector with matter by a generic k-essence field describing two kinetic regimes: A standard Klein-Gordon Lagrangian in the lower bound and a Cuscuton action in the opposite extremal case. In addition to this, we find a specific k-essence field which has appropriate limits on both kinetic bounds. The Chameleon effect is a key component of our discussion, since the coupling to matter generates an effective potential with properly defined equilibrium points. \\ In Section \ref{sectfour}, we use this model to analyze the Early Universe Cosmology from a perspective different from inflation. With real expectations on the possibilities to reproduce Particle Physics at low energies.\\ In the final section we discuss the results and conclude.
\newpage
\section{The Induced Metric}\la{secttwo}
In this section we introduce an induced geometry to describe the dynamics in the matter sector. The concepts and tools of Finsler Geometry have been intensively mentioned in the literature \cite{Bekenstein:1992pj}, \cite{Girelli:2006fw} since these respond to our call for a change in the dispersion relations obeyed by matter. This technology is crucial in order to connect each modified dispersion relation with an induced metric. 
\subsection{Preliminaries}\la{secttwoone}
In \cite{Chern:1996mt} we find a concise definition of a Finsler Geometry:
\begin{flushright}
\par \emph{Finsler Geometry Is Just Riemannian Geometry without the Quadratic Restriction}\\
\emph{S.S. Chern}
\end{flushright}
Originally, Riemann was the first to approach these concepts as natural extensions of his own work. But the name ``Finsler Geometry'' came from Finsler's thesis in Gottingen in 1918.\\  
Meanwhile we should ask: what is this good for? The answer for this question relies on the fact that most of the kinematic properties we know depend on standard dispersion relations. In \cite{Girelli:2006fw}, we learn that it is possible to relate a modified dispersion relation with(out) non-quadratic terms with an induced metric. And it is convenient in order to consider a wider spectrum of new possibilities to study particle dynamics.\\ 
To see this, we write the arc length:
\be
I=m\int\limits_a^b F(x,\dot{x})d\tau
\la{linel}
\ee
With $\dot{x}=dx/d\tau$. At first sight, we identify the Finsler function $F$ to be ``velocity dependant''. Then, the metric is defined by:
\be
g_{\alpha\beta}(x,\dot{x})\equiv\frac{1}{2}\frac{\partial^2 F^2}{\partial \dot{x}^\alpha \partial \dot{x}^\beta}
\la{finmetric}
\ee
And the inverse is defined by $g_{\alpha\beta}(x,\dot{x})g^{\alpha\gamma}(x,\dot{x})=\delta_\beta^\gamma$.\\
As a consistency check, we can use $F=\sqrt{g_{\alpha\beta}(x)\dot{x}^\alpha\dot{x}^\beta}$ when $g_{\alpha\beta}$ only depends on the spacetime coordinates to recover the components of the metric tensor as usual.
\newpage
From this check, we notice a curious property of the $F$ function:
\be
F(x,\lambda\dot{x})=|\lambda|F(x,\dot{x})
\la{homog}
\ee
This means that $F$ is an homogeneous function of first degree, which implies the independence of velocities and spacetime coordinates. In such a case the Euler's Theorem states:
\be
\dot{x}^\alpha\frac{\partial F(x,\dot{x})}{\partial \dot{x}^\alpha}=F(x,\dot{x})
\la{eulerF}
\ee
Using [\ref{finmetric}] and [\ref{eulerF}] we find a general solution for $F$:
\be
F=\sqrt{g_{\alpha\beta}(x,\dot{x})\dot{x}^\alpha\dot{x}^\beta}
\la{Fgen}
\ee
From [\ref{Fgen}], we notice that if $F$ is an homogeneous function of first degree, then $g_{\alpha\beta}(x,\lambda\dot{x})=g_{\alpha\beta}(x,\dot{x})$. This means that the degree of homogeneity for the metric is zero.\\ 
The equivalent to [\ref{eulerF}] for the metric reads:
\be
\dot{x}^\alpha\frac{\partial g_{\gamma\beta}(x,\dot{x})}{\partial \dot{x}^\alpha}=0
\la{eulerg}
\ee
Using this fact, the variations of the action [\ref{linel}] with respect to the velocities do not contribute at all. And the geodesic equation is:
\be
\ddot{x}^\alpha+\Gamma^\alpha_{\beta\gamma}(x,\dot{x})\dot{x}^\beta\dot{x}^\gamma=0
\la{geodfins}
\ee
The Christoffel symbols are:
\be
\hat{\Gamma}^\alpha_{\beta\gamma}(x,\dot{x})=\frac{g^{\alpha\kappa}(x,\dot{x})}{2}\left( g_{\beta\kappa}(x,\dot{x})_{,\alpha}+ g_{\alpha\kappa}(x,\dot{x})_{,\beta}-g_{\beta\alpha}(x,\dot{x})_{,\kappa}\right)
\la{chrisfins}
\ee 
And in the same way, the geodesic equation can be rewritten in terms of a covariant derivative:
\ba
\hat{\nabla}_\mu v^\nu&\equiv&v^\nu_{,\mu}+\hat{\Gamma}^\nu_{\alpha\mu}(x,\dot{x})v^\alpha\nonumber\\
v^\alpha\hat{\nabla}_\alpha v^\beta&=&0
\la{Fincovdev}
\ea 
With $\phi_{,\alpha}\equiv\partial\phi/\partial x^\alpha$. These are not so different from the expressions we normally use in GR\footnote{In \cite{Kouretsis:2012ys} we can find a nice revision of other geometrical entities in Finsler Geometries relevant to be compared.}, the difference comes from the velocity dependence of the metric.\\ Now, our task reduces to find $F$ for a given on-shell dispersion relation $\mathcal{M}(p)=m^2$.\footnote{The extension we consider is still metric, non-metric structures are not part of this analysis.}
We must define the conjugate momentum as follows:
\be
p_\mu=m\frac{\partial F}{\partial \dot{x}^\mu}=m\frac{g_{\mu\nu}(x,\dot{x})\dot{x}^\nu}{F}
\la{mom}
\ee
Using this definition, we notice that $H=\dot{x}^\mu p_\mu-\sqrt{g_{\alpha\beta}(x,\dot{x})\dot{x}^\alpha \dot{x}^\beta}=0$. Thus we are forced to introduce a Lagrange multiplier ($\kappa$):
\be
H=\dot{x}^\mu p_\mu-\kappa\left(\mathcal{M}(p)-m^2\right)
\la{Hamilton} 
\ee
From the last expression, first Hamilton's equation reads:
\be
\dot{x}^\alpha=\kappa\frac{\partial \mathcal{M}(p)}{\partial p_\alpha}
\la{xpunto}
\ee
Inverting these equations we find $p^\alpha=f(\dot{x}^\beta,\kappa)$, and using the inverse Legendre transformations we find the Lagrangian. With the equations of motion for $\kappa$ the Lagrangian becomes an expression purely dependant on coordinates and velocities. Thus, the action can be written as:
\be
I=\int \mathcal{L}(x,\dot{x})d\tau
\nonumber
\ee 
Comparing with [\ref{linel}] we find:
\be
\mathcal{L}(x,\dot{x})=mF(x,\dot{x})
\la{finalpre}
\ee
The last result allows us to find a Finsler function $F$ (and a metric by using [\ref{finmetric}]) for a given a dispersion relation. Let us emphasize that in the case in which the modified dispersion relation does not depend on aquadratic terms, we just need [\ref{finmetric}] to find an induced metric. However, we will test the connection between Finsler functions and modified dispersion relations in the ``worst-case scenario''. To do so, we present an example to clarify this procedure. Let us consider the following 1-D dispersion relation with an aquadratic term:
\be
\mathcal{M}(p)=-p_0^2\left(1+\alpha\frac{p_0}{M}\right)+p_1^2
\la{Mex}
\ee
Now we write the corresponding action as a function of the Lagrange multiplier $\kappa$ 
\be
I=\int d\tau\left(\dot{t}p_0+\dot{x}p_1-\kappa\left(\mathcal{M}(p)-m^2\right)\right)
\la{exact}
\ee
Varying the action with respect to $p_\mu$, we get:
\ba
\dot{t}&=&-\kappa\left(2p_0+3\frac{\alpha}{M}p_0^2\right)\nonumber\\
\dot{x}&=&2\kappa p_1\la{Hamex}
\ea
Hence, the conjugate momenta are:
\ba
p_0&=&\frac{M}{3\alpha}\left[-1+\sqrt{1-\frac{3\alpha}{M\kappa}\dot{t}}\right]\la{spmom}\\
p_1&=&\frac{\dot{x}}{2\kappa}\nonumber
\ea
Where we picked a regular solution in the limit $\alpha\rightarrow 0$. Replacing in [\ref{exact}] we find:
\be
I=\int d\tau\left(\frac{-\dot{t}^2+\dot{x}^2}{4\kappa}+\frac{\alpha\dot{t}^3}{8\kappa^2 M}+\kappa m^2+O(\alpha^2)\right)
\la{LEx}
\ee
The variation of [\ref{LEx}] respect to the Lagrange multiplier gives an approximate solution for $\kappa$:
\be
\kappa(\dot{t},\dot{x})\approxeq \frac{\sqrt{-\dot{t}^2+\dot{x}^2}}{2m}+\frac{\alpha}{2M}\frac{\dot{t}^3}{-\dot{t}^2+\dot{x}^2}
\nonumber
\ee
The full Lagrangian now reads:
\be
\mathcal{L}(x,\dot{x})=m\sqrt{-\dot{t}^2+\dot{x}^2}+\frac{\alpha m^2}{2M}\frac{\dot{t}^3}{-\dot{t}^2+\dot{x}^2}\nonumber
\ee
By [\ref{finalpre}], the Finsler function is:
\be
F(x,\dot{x})=\sqrt{-\dot{t}^2+\dot{x}^2}+\frac{\alpha m}{2M}\frac{\dot{t}^3}{-\dot{t}^2+\dot{x}^2}\nonumber
\ee
And finally [\ref{finmetric}] provides the metric components at first order in $\alpha$\footnote{Being rigorous, the Finsler function fails to fulfill [\ref{homog}] for negative values of $\lambda$. These cases are considered by \cite{Girelli:2006fw} as \textit{positively homogeneous}.}:
\ba
g_{00}(x,\dot{x})&=&-1+\left(\frac{\alpha m\dot{t}}{2M}\right)\frac{2\dot{t}^4+6\dot{x}^4-5\dot{x}^2\dot{t}^2}{\left(-\dot{t}^2+\dot{x}^2\right)^{5/2}}\nonumber\\
g_{11}(x,\dot{x})&=&1+\left(\frac{\alpha m\dot{t}^3}{M}\right)\frac{\dot{t}^6-3\dot{x}^4\dot{x}^4\dot{t}^2+2\dot{t}^6}{\left(-\dot{t}^2+\dot{x}^2\right)^{9/2}}\nonumber\\
g_{10}(x,\dot{x})&=&g_{01}(x,\dot{x})=-\frac{3\alpha m\dot{t}^2\dot{x}^3}{M\left(-\dot{t}^2+\dot{x}^2\right)^{5/2}}\la{metex}
\ea
In the limit $\alpha\rightarrow 0$, these are the components of the Minkowski metric with signature $(-+++)$, which will be followed from now on. 
\newpage
\subsection{Constraining the model}\la{secttwotwo}
In the previous subsection, we learned to identify each dispersion relation with a Finsler function. Knowing this, our interest sets into a specific set of Finsler functions to describe the dynamics of the matter sector.\\ From now on, we follow \cite{Bekenstein:1992pj} in a more conservative perspective than the one we developed in the last example. We will use this technology to aim for a modified dynamics without non-quadratic terms.\\
Considering [\ref{Fgen}], we may rewrite the differential line element in [\ref{linel}] as follows:
\be
ds^2=g_{\alpha\beta}(x)\dot{x}^\alpha \dot{x}^\beta  G(x,dx^1/dx^0,dx^2/dx^0,dx^3/dx^0)d\tau^2
\la{genFins}
\ee
This is still the most general way to write the line element by identifying $g_{\alpha\beta}(x)\dot{x}^\alpha \dot{x}^\beta G\equiv F^2(x,\dot{x})$, and $m=1$. The difference relies on $G(x,dx^i/dx^0)$, which carries all possible powers of $\dot{x}$  and is an homogeneous function of degree zero with respect to them. The function is written in terms of the frame dependent ratios $dx^i/dx^0$ to keep the degree of homogeneity. $g_{\alpha\beta}$ is just the standard gravitational metric. But there is something wrong with this expression: it becomes extremely hard to build a coordinate invariant $G$ out of three independent variables. And the fact that we just count on three (and not four) of them implies a violation of the spirit of covariance for the theory.\\
To solve these issues, we draw our attention to a metric we can write from [\ref{genFins}] by using [\ref{finmetric}] {\bf and does not depend on the velocities}. Moreover, we are obliged to write $G$ in terms of coordinate invariants. So far, we have written the only choice we have: $g_{\alpha\beta}\dot{x}^\alpha \dot{x}^\beta$. \footnote{Excluding $\epsilon_{\alpha \beta \gamma \delta}\dot{x}^\alpha \dot{x}^\beta \dot{x}^\gamma \dot{x}^\delta$ to hold linear equations of motion.}  We need additional fields with contracting vector indices to write more of these expressions:
\ba
X_\phi&=&-\frac{1}{2}g^{\alpha\beta}\phi_{,\alpha} \phi_{,\beta}\nonumber\\
H&=&-\frac{\left(\phi_{,\mu} \dot{x}^\mu\right)^2}{g_{\alpha\beta}\dot{x}^\alpha \dot{x}^\beta}\la{scinv}
\ea
Another option is to use a vector field $A_\alpha$ in the same fashion. Both are invariant quantities with degree zero of homogeneity and depending on {\bf four} velocities. The introduction of more degrees of freedom is not logically excluded, such an addition is not considered just to preclude the theory from higher order contributions.
\newpage
Now the line element built from invariants is:
\be
ds^2=g_{\alpha\beta}\dot{x}^\alpha \dot{x}^\beta G(X_\phi,H,\phi)d\tau^2 
\la{lintwo}
\ee 
After these considerations for the Finsler function, we use [\ref{finmetric}] to find the metric
\ba
\tilde{g}_{\alpha\beta}&=&\frac{1}{2}\frac{\partial^2 F^2}{\partial \dot{x}^\alpha \partial \dot{x}^\beta}\nonumber\\
&=&\left(G-HG'\right)g_{\alpha\beta}-\left(G'+2HG''\right)\phi_{,\alpha} \phi_{,\beta}-2H^2G''\left[\frac{\phi_{,(\alpha}g_{\beta\mu)}\dot{x}^\mu}{\phi_\nu\dot{x}^\nu}-\frac{\dot{x}_\alpha\dot{x}_\beta}{\dot{x}_\alpha\dot{x}^\alpha}\right]\nonumber
\ea
where $G'\equiv dG/dH$ and $\dot{x}^\alpha\equiv g^{\alpha\beta}\dot{x}_\beta$. If $G$ is a linear function in $H$, then $\tilde{g}_{\alpha\beta}$ is completely independent from $\dot{x}^\alpha$:
\be
G(X_\phi,H,\phi)=A(X_\phi,\phi)+B(X_\phi,\phi)H
\la{Gfin}
\ee
By replacing in [\ref{lintwo}] we generate an induced metric:
\be
\tilde{g}_{\alpha\beta}=A(X_\phi,\phi)g_{\alpha\beta}-B(X_\phi,\phi)\phi_{,\alpha}\phi_{,\beta}
\la{indmet}
\ee
The relation is analogous when we pick a vector to generate the invariants. Not surprisingly, the metric is just modified by the symmetric product of the extra degrees of freedom. From [\ref{indmet}], notice that the relation between $\tilde{g}_{\alpha\beta}$ and $g_{\alpha\beta}$ is not necessarily conformal. Writing the corresponding expression for the inverse metric:
\be
\tilde{g}^{\alpha\beta}=A^{-1}(X_\phi,\phi)\left[g^{\alpha\beta}+\frac{B(X_\phi,\phi)}{A(X_\phi,\phi)+2B(X_\phi,\phi)X_\phi}\phi^{,\alpha}\phi^{,\beta}\right]
\la{invindmet}
\ee
With $\phi^{,\alpha}\equiv g^{\alpha\beta}\phi_{,\beta}$. 
Extensive studies have been performed when $B=0$, in the so-called Brans-Dicke Gravity (and its cosmological implications in \cite{Steinhardt:1989prl}).
Prescribing that $g_{\mu\nu}\dot{x}^\mu \dot{x}^\nu=0$ is not a solution of the equation
\be
F(X_\phi,H,\phi)g_{\mu\nu}\dot{x}^\mu \dot{x}^\nu=0
\nonumber
\ee
means that a surface which is null for gravitons is not for other species.\\
Our interest is focused in the disformal version of [\ref{indmet}], and explicitly when $B>0, A> 0$. Recalling our choice for the signature, we evaluate the null condition for the new metric: 
\be
0=\tilde{g}_{\mu\nu}v^\mu v^\nu=A(X_\phi,\phi)g_{\mu\nu}v^\mu v^\nu-B(X_\phi,\phi)(\phi_{,\alpha}v^\alpha)^2
\nonumber
\ee
Which implies:
\be
g_{\mu\nu}v^\mu v^\nu=\frac{B}{A}(\phi_{,\alpha}v^\alpha)^2\geq 0
\la{suplum}
\ee
For $B,A>0$, according to the induced metric the vector $v$ is null, but it is spacelike in $g$. 
This means that the speed of light defined in the matter sector (for particles using $\tilde{g}_{\mu\nu}$ to describe dynamics) is faster than ``the speed of light'' ruling the gravitational sector. The following picture illustrates our point: 
\begin{center}
\begin{figure}[h]
\centering
\includegraphics[width=5cm, height=5cm]{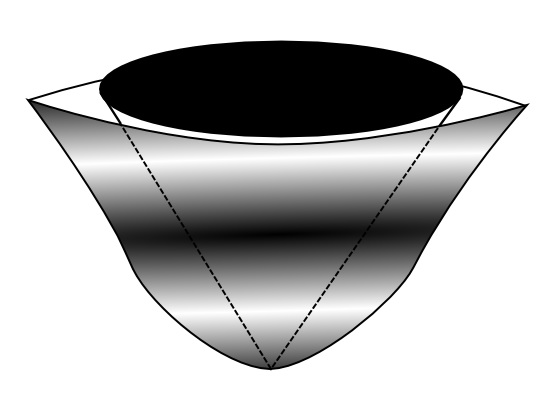}
\caption{\small{Two features of the lightcone generated by the induced metric $(B>0,A\geq 0)$: (a) It is wider that the original considered in the gravitational sector. This is a hint to solve the horizon problem in cosmology. (b) It is wiggled to represent the variation of speed of light with time. Space homogeneity can be easily removed by imposing anisotropic solutions for $\phi$.}}
\label{fig1}
\end{figure}
\end{center}

Highlighting the fact that a different choice of signs for $A$ and $B$ might not lead us to the same results. For example, Modified Newtonian Dynamics (MOND) theories were not conceived originally to be superluminal \cite{Bekenstein:1992pj}. And for instance, $A$ and $B$ have different signs.\\
The idea of an enlarged causal area is a crucial point in this model. But if our understanding is based solely in our notions of GR, we may find reasonable objections related to noncausal propagation of superluminal degrees of freedom.\\
\newpage
However, the arguments used in [\cite{Bruneton:2006gf},\cite{Babichev:2007dw}] are useful in order to clarify some of these concerns. Here we have a classical paradox as an example:

\begin{figure}[h]
\centering
\includegraphics[width=12cm, height=7cm]{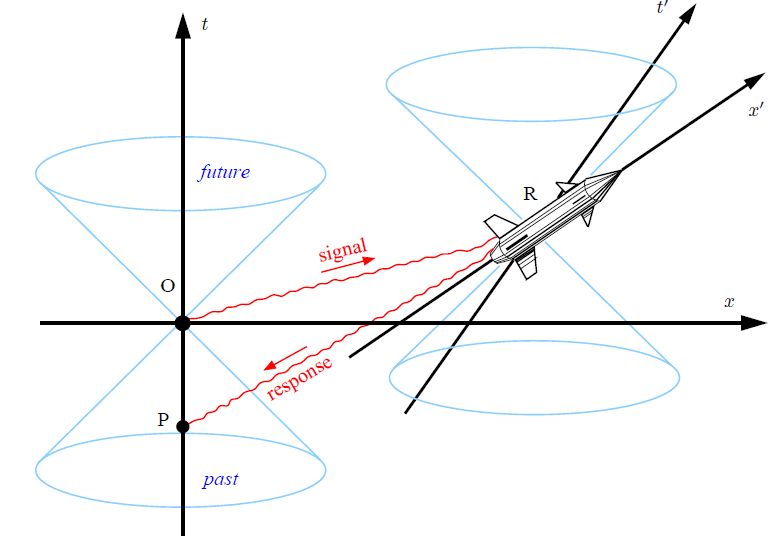}
\caption{\small{A superluminal signal sent to a moving spaceship: The response arrives {\bf before} the signal was emmited}}
\label{fig2}
\end{figure}
This figure makes reference to the well-known ``tachyonic antitelephone'' in which an observer at rest sends a signal with an arbitrarily large speed to a moving spaceship, then someone inside the vehicle responds the signal. If the velocity of the spacecraft is large enough, the returning signal is received by the observer at rest before it was emitted. The final result is that the observer at rest sent a signal to its own past.
\newpage
In the next figure (See Figure [\ref{fig2}]), we see a way in which this conundrum is remediated is by considering a deformed lightcone just a bit simpler than the one we depicted in Figure [\ref{fig1}], in which the signal is still sent with a speed larger than $c$. But it is now received in the future, with no causal violations.\\
\begin{figure}[h]
\centering
\includegraphics[width=11cm, height=6.5cm]{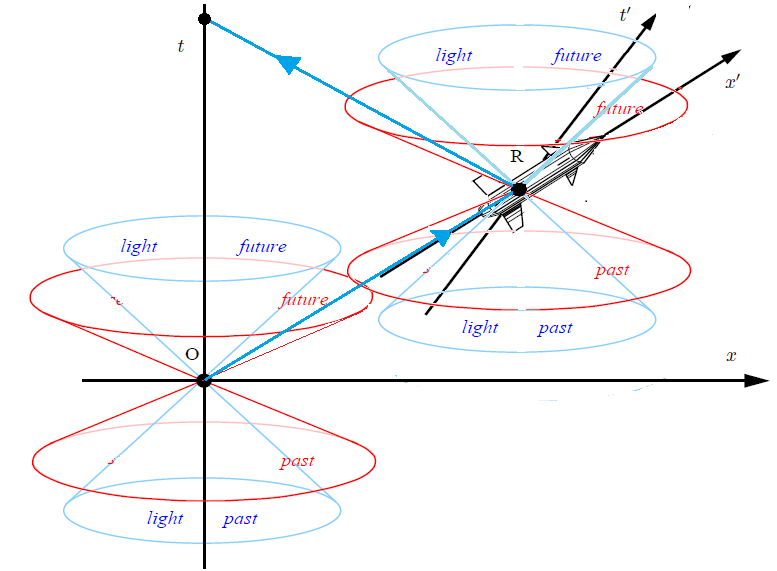}
\caption{\small{The paradox is avoided by considering a wider lightcone (in a simplest case, generated by a nearly constant field), signals are emitted by the observer in the spaceship, and received in the future by the observer at rest. The emmited signal is inside the extended future lightcone for $O$, and the received signal is inside the extended past lightcone of $R$. When the signal is retransmitted, it is emmited in the extended future lightcone of $R$ and is directed towards the causal future of $O$.}}
\label{fig3}
\end{figure}

In \cite{Babichev:2007dw}, the authors refer to the scalar field used in the induction as an \emph{aether} picking the ``right'' lightcone. The geometric construction we made in this section makes reference to an additional degree of freedom useful to reparametrize. From now on, we refer to it as a ``gravitational scalar'' (or vector, depending on the choice made to generate the invariants in [\ref{scinv}]).
\newpage
\subsection{Other induction method}\la{secttwothree}
In addition to the induction technique described in previous parts, let us compare our results with another alternative in the existing literature \cite{Babichev:2007dw}. Consider the k-essence field action:
\ba
I_\phi&\equiv&\int d^4 x\sqrt{-g}\mathcal{L}(X,\phi)\nonumber\\
X&\equiv& -\frac{1}{2}g^{\mu\nu}\nabla_\mu\phi\nabla_\nu\phi\la{defoth}
\ea
Where $g_{\mu\nu}$ is the gravitational metric. Comparing this to the action of a perfect fluid, we define:
\ba
\rho&=&2X\mathcal{L}_{,X}-\mathcal{L}\nonumber\\
p&=&\mathcal{L},\la{defsc}
\ea
where $\mathcal{L}$ does not depend on higher derivatives of the inducing field. There is nothing misterious about these definitions, such expressions make perfect sense in the case $\mathcal{L}=X-V$ which is just the usual analogy between the energy-momentum tensor for a fluid and a canonical scalar field. And hence we can associate a \emph{speed of sound}:
\be
c_s^2=\frac{\partial p}{\partial \rho}=\frac{p_{,X}}{\rho_{,X}}=\left(1+2X\frac{\mathcal{L}_{,XX}}{\mathcal{L}_{,X}}\right)^{-1}
\la{cs}
\ee
Varying this action with respect to the field we find the equations of motion:
\ba
\frac{\delta I_\phi}{\delta\phi}&=&G^{\mu\nu}\nabla_\mu\nabla_\nu\phi-2X\mathcal{L}_{,X\phi}+\mathcal{L}_{,\phi}\nonumber\\
G^{\mu\nu}&\equiv&\mathcal{L}_{,X}g^{\mu\nu}-\mathcal{L}_{,XX}\nabla^\mu\phi\nabla^\nu\phi\la{defothmet}
\ea
In here we notice the presence of $G^{\mu\nu}$ as an induced metric. Which in the case $\mathcal{L}=X-V$ just returns $G^{\mu\nu}=g^{\mu\nu}$. Now, the inverse metric is:
\be
G_{\mu\nu}=\mathcal{L}_{,X}^{-1}\left(g_{\mu\nu}+\frac{\mathcal{L}_{,XX}}{\mathcal{L}_{,X}+2X\mathcal{L}_{,XX}}\nabla_\mu\phi\nabla_\nu\phi\right)
\la{indoth}
\ee
Where:
\ba
T_{\mu\nu}^{(\phi)}n^\mu n^\nu>0\rightarrow\mathcal{L}_{,X}>0 &\longrightarrow& \text{(Null energy condition)}\nonumber\\
1+2X\frac{\mathcal{L}_{,XX}}{\mathcal{L}_{,X}}>0&\longrightarrow& \text{(Hyperbolicity condition for [\ref{defothmet}])\nonumber}
\ea
Thus, by recalling [\ref{indmet}] we can match the conditions $B,A>0$ assumed in order to have an expanded lightcone in [\ref{suplum}] with [\ref{indoth}], we find:
\ba
A>0&\longrightarrow&\mathcal{L}_{,X}>0\la{ALX}\\
B>0&\longrightarrow&1+2X\frac{\mathcal{L}_{,XX}}{\mathcal{L}_{,X}}>0, \:\:\mathcal{L}_{,XX}<0\la{BLX}
\ea
This comparison not only has shown an obvious analogy with our previous results; but also it provides consistence with our assumptions of superluminosity. As a method of induction, it shares many similarities with the example used in the first subsection. In [\ref{ALX}], the prescription $A>0$ agrees with the null energy condition, which prevents us from energy values unbounded below. And [\ref{BLX}] allows us to reproduce the so-called hyperbolicity condition to solve the equations of motion in \ref{defothmet}. In the context of fluid mechanics, the same statement allows the propagation of ``sound waves'' with $c_s^2$ being positive. \\
In the Appendix \ref{appone}, we can find a suitable description of the dynamics for the perturbations of $\phi$ in a geometry conformally related with $G^{\mu\nu}$.\\Also, we should notice that a generalization of the k-essence field containing higher derivative contributions is consistent with the description in Section \ref{secttwoone}. Such a case will not be discussed in the rest of this dissertation, but it is sensible in the context of more profound modifications.\\\\
At this stage, it is important to mention that this is just a good motivation to find a suitable mapping of possible inductions of k-essence fields into bimetric theories. We cannot neglect that both modified dispersion relations and generalized k-essence lagrangian densities have been used as generating functions of these induced geometries. Later in this manuscript, we describe an alternative to identify certain k-essence theories which lead us into the simplest disformal bimetric models.
\newpage
\section{Action and motion equations}\la{sectthree}
\subsection{General structure and matter coupling conditions}\la{sectthreeone}
Our study focuses in a generic action as the one that follows:
\ba
I&=&I_g+I_\phi+I_m\la{genact}\\
&=&\int d^4x\sqrt{-g}R(g_{\mu\nu})+\int d^4x\sqrt{-g}\mathcal{L}_\phi(X,\phi)+\int d^4x\sqrt{-\tilde{g}}\left[\mathcal{L}_m(\Psi_m,\tilde{g}_{\mu\nu})\right]
\nonumber
\ea
Where $R(g_{\mu\nu})$ is the Ricci scalar with the classical aspect it inherits from GR. $\mathcal{L}_\phi(X,\phi)$ is the k-essence Lagrangian for the gravitational scalar\footnote{This field is also called a ``bi-scalar'' in the literature, since it transforms like a scalar in the two frames.} analogous to the expressed in [\ref{defoth}. And $\mathcal{L}_m(\Psi_m,\tilde{g}_{\mu\nu})$ is the matter Lagrangian, which purely depends on the induced metric [\ref{indmet}]. In principle, the construction developed in Section [\ref{secttwo}] is not imposing any specific choice of $\mathcal{L}_\phi$. On the other hand, in addition to all other rigid and local symmetries in the matter sector, the presence of a different lightcone lead us to think about the symmetries in [\ref{genact}]. And following the spirit of GR, we modify the preexisting Lorentz symmetry by doubling it:
\ba
g_{\mu\nu} &=& g_{\alpha\beta}\Lambda^\alpha_\mu\Lambda^\beta_\nu\nonumber\\
\tilde{g}_{\mu\nu} &=& \tilde{g}_{\alpha\beta}L^\alpha_\mu L^\beta_\nu\la{Lmod}
\ea 
The modification responds minimally to what is described in Figure [\ref{fig3}]: the change in the speed of light defines another scale for the parameters involved in the transformations of objects with Lorentz indices in the matter sector:
\ba
\Lambda^\alpha_\mu&=&\exp[i\theta^k \left(\mathcal{T}^k\right)^\alpha_\mu]\nonumber\\
L^\alpha_\mu&=&\exp[i\beta(x^\mu)\theta^k \left(\mathcal{T}^k\right)^\alpha_\mu]\la{modlorentz}
\ea
Where $\left(\mathcal{T}^k\right)^\alpha_\mu$ are the generators of the Lorentz algebra $\mathbf{so}(3,1)$, these are the same for the two transformations. For the scale $\beta(x^\mu)$, a Taylor expansion shows that the transformations of spacetime dependent vectors remain linear just in a region around a point. Moreover, observing the lightcones in Figure [\ref{fig1}], our intuition suggests an explicit relation between $\beta$ and the inducting field $\phi$.\\An example of this expression will be described later in the discussions.\\
In the previous sections, we achieved a brief description of a geometry for the matter sector with an expanded lightcone. We proceed with our endeavours by asking if ordinary matter fields can propagate superluminally. To give a justified answer, we are motivated by the discussions in \cite{Afshordi:2006ad}. In here, the authors analyze the interaction of a k-essence field called {\it Cuscuton} with a massive scalar field in a Minkowski background. Our calculation shares essentially the same spirit: it is a dynamical sketch of the way in which the perturbations of the matter field propagate. But it certainly differs in the application of the induced geometry for the matter sector. For simplicity, we have considered the geometry as emergent from a flat Minkowski background.\\
In this case, the action for the system is\footnote{Warning: we suggest the reader to be extremely careful about the metric used to raise and lower indices, since $\tilde{g}_{\mu\nu}A^\mu\neq g_{\mu\nu}A^\mu$, and in a similar way for the contravariant components. The transformation laws between these two objects are found just by direct application of the induced metric on a vector (or covector). So far, we have only used the gravitational metric to raise and lower indices. We will explicitly mention the cases in which we need [\ref{indmet}] and [\ref{invindmet}] to do so.}
\be
I=\int d^4x\left(\mu^2\sqrt{2X_\phi}-\frac{1}{2}m_\phi^2\phi^2\right)+\int d^4x \sqrt{-\tilde{\eta}}\left[-\frac{1}{2}\tilde{\eta}^{\alpha\beta}\psi_{,\alpha}\psi_{,\beta}-\frac{1}{2}m_\psi^2\psi^2\right]
\la{actsupex}
\ee  
By neglecting all conformal modificacions $(A=1)$, we use [\ref{invindmet}] to build an induced metric
\ba
\tilde{\eta}^{\alpha\beta}&\approx&\eta^{\alpha\beta}+B\phi^{,\alpha}\phi^{,\beta}\nonumber\\
det(\tilde{\eta}_{\alpha\beta})&=&1+2BX_\phi\la{metdet}
\ea
Where $X_\phi=-1/2\:\eta^{\alpha\beta}\phi_{,\alpha}\phi_{,\beta}$ and $X_\psi=-1/2\:\eta^{\alpha\beta}\psi_{,\alpha}\psi_{,\beta}$.\\ And thus, we write the full action with respect to the Einstein frame, and by quoting this we mean that is written in terms of the original metric (in this case $\eta_{\alpha\beta}$). Keeping only first order terms in $B$, we find:
\be
I=\int d^4x\left[\mu^2\sqrt{2X_\phi}+X_\psi+BX_\phi X_\psi-\frac{1}{2}B\left(\eta^{\alpha\beta}\phi_{,\alpha}\psi_{,\beta}\right)^2-\frac{1}{2}m_\psi^2\psi^2-\frac{1}{2}m_\phi^2\phi^2\right]
\la{fullcouplact}
\ee 
Now we draw our attention to the last three terms in the action, if we decompose the fields as follows:
\ba
\phi(x,t)&=&\phi(t)+\delta\phi(x,t)\nonumber\\
\psi(x,t)&=&\psi(t)+\delta\psi(x,t)\nonumber
\ea
At the lowest nontrivial order in perturbations, those terms are combined in a single expression:
\be
V(\delta\phi,\delta\psi)=-2B\dot{\phi}^2X_{\delta\psi}-2B\dot{\psi}^2X_{\delta\phi}-\frac{B}{2}\dot{\psi}\dot{\phi}\delta\psi^{,\alpha}\delta\phi_{,\alpha}+\frac{1}{2}m_\psi^2\delta\psi^2+\frac{1}{2}m_\phi^2\delta\phi^2\nonumber
\ee
Considering the Fourier transformed perturbations and $\left(\mathbf{k}^2-\omega^2\right)\sim\mathbf{k}^2$, rearranging the terms of the last expression:
\ba
V(\delta\phi_\mathbf{k},\delta\psi_\mathbf{k})&=&-B\mathbf{k}^2\dot{\phi}^2(\delta\psi_\mathbf{k})^2-B\mathbf{k}^2\dot{\psi}^2(\delta\phi_\mathbf{k})^2+\frac{B}{2}\mathbf{k}^2\dot{\psi}\dot{\phi}\delta\psi_\mathbf{k}\delta\phi_\mathbf{k}\nonumber\\
&+&\frac{1}{2}m_\psi^2(\delta\psi_\mathbf{k})^2+\frac{1}{2}m_\phi^2(\delta\phi_\mathbf{k})^2\nonumber\\
&=&\frac{V_{,\phi\phi}}{2}\delta\phi_\mathbf{k}^2+\frac{V_{,\psi\psi}}{2}\delta\psi_\mathbf{k}^2+V_{,\phi\psi}\delta\phi_\mathbf{k}\delta\psi_\mathbf{k}.
\la{potential}
\ea
Thus, the action is now 
\be
I=\int d^4x\left[\mu^2\sqrt{2X_\phi}+X_\psi-V(\phi,\psi)\right]
\la{actmodghaz}
\ee
And we just simply go through the calculations made by the authors in \cite{Afshordi:2006ad}, writing the perturbations for the Cuscuton and the matter field:
\ba
\delta\phi_\mathbf{k}&=&-\left(\frac{V_{,\phi\psi}}{k^2+V_{,\phi\phi}}\right)\delta\phi_\mathbf{k}\nonumber\\
(\omega^2-k^2)\delta\psi_\mathbf{k}&=&V_{,\phi\psi}\delta\phi_\mathbf{k}+V_{,\psi\psi}\delta\psi_\mathbf{k}\nonumber
\ea
The dispersion relation is:
\be
\omega^2=\frac{k^4+(V_{,\phi\phi}+V_{,\psi\psi})k^2+V_{,\phi\phi}V_{,\psi\psi}-V_{,\phi\psi}^2}{k^2+V_{,\phi\phi}}
\la{disprelsimp}
\ee
In the short-wavelength regime, we can verify that:
\be
v_g=\frac{d\omega}{dk}=1-\frac{V_{,\psi\psi}}{2k^2}+O\left(\frac{V_{,\phi\psi}^2}{k^4}\right)
\la{moddr}
\ee
By replacing $V_{,\psi\psi}$ from [\ref{potential}] we find:
\be
v_g=1+B\dot{\phi}^2-\frac{m_\psi^2}{2k^2}+O\left(\dot{\phi}^4,\dot{\phi}^2\dot{\psi}^2\right)
\la{supmoddr}
\ee
Superluminal dispersion ($v_g>1$) demands dominance of the coupling over the mass term.\\ So now this nice example gives us the basis we needed to put superluminal propagation of fields under the spotlight: it relies on these \emph{tachyonic} coupling terms, and is precisely the condition $B>0$ which provides that unique feature. The tachyonic terms in the potential arise many concerns related to instability, this is a sensible concern at the interaction level in the matter sector, and at this moment is when we must remember that this procedure follows in the Einstein frame.  We can also argue that the existence of such a minima that allows a controlled expansion is controversial. However, seen in this frame or in the matter frame, the presence of (\emph{meta})stable configurations of the potential is a mild requirement for the validity of this last statement. In the next section we clarify some ideas related to this issue.\\ The argument developed lines above works straightforwardly for a canonical Klein-Gordon Lagrangian density instead of a Cuscuton. And in fact, the result holds for any k-essence field in which the kinetic term is proportional to $k^2\delta\phi_{\mathbf{k}}^2$ at first order in perturbations.
\subsection{Exploring the basics: $\mathcal{L}_\phi=X-V$}\la{sectthreetwo}
Based in \cite{Clayton:1999zs}, \cite{Noller:2012sv}, here we consider a standard Klein-Gordon Lagrangian density ($\mathcal{L}_\phi=X_\phi-V(\phi)$) as an option to explore the dynamics of the gravitational scalar used to induce the matter geometry. We are specially interested in this case since it is a prudent lower limit for a generic k-essence action. Hence, the appropriate expression for [\ref{genact}] is:
\be
I\equiv I_E+I_m=\int d^4x\sqrt{-g}\left[R(g_{\mu\nu})+X_\phi-V(\phi)\right]+\int d^4x\sqrt{-\tilde{g}}\left[\mathcal{L}_m(\Psi_m,\tilde{g}_{\mu\nu})\right].
\la{KGact}
\ee
If we recall [\ref{indmet}], there is an explicit dependence of $A$ and $B$ in the field and its derivatives. This is why the results obtained by Noller in \cite{Noller:2012sv} are relevant in our discussion.\\
In addition to the symmetries mentioned in the last subsection [\ref{sectthreeone}], the fields in the matter sector preserve the action invariant under traslations in the induced geometry
\be
\tilde{\nabla}_\mu(\sqrt{-\tilde{g}}\tilde{T}^{\mu\nu})=0
\la{trinv}
\ee
Where $\tilde{\nabla}$ is the covariant derivative in the induced metric. The conserved current can be found as:
\be
\tilde{T}^{\mu\nu}=\frac{2}{\sqrt{-\tilde{g}}}\frac{\delta I_m}{\delta \tilde{g}_{\mu\nu}}
\la{defT}
\ee
The conserved energy-momentum tensor in [\ref{trinv}] does not include any contributions from $I_\phi$. If we decide to map all the expressions in the matter action to the Einstein frame and define:
\be
T^{\mu\nu}=\frac{2}{\sqrt{-g}}\frac{\delta}{\delta g_{\mu\nu}}(I_m+I_\phi)
\la{defTEin}
\ee
We see that by varying the total action with respect to the gravitational metric we recover Einstein's equations. Bianchi identities are trivially fulfilled. But if we stay with the definition given in [\ref{defT}] and include the action for the scalar field we find $G_{\mu\nu}=8\pi G\sqrt{\tilde{g}/g}\:\tilde{T}_{\mu\nu}$. The consistency Bianchi identities is not a trivial statement to proof.\\
Now, we calculate the equations of motion for the inducing field by using the chain rule:
\be
\frac{\delta I}{\delta\phi}=\frac{\delta I_E}{\delta\phi}+\frac{\delta I_m}{\delta\tilde{g}_{\mu\nu}}\frac{\delta \tilde{g}_{\mu\nu}}{\delta\phi}=\frac{\delta}{\delta\phi}(X_\phi-V(\phi))+\frac{\delta \tilde{g}_{\mu\nu}}{\delta\phi}\frac{\delta}{\delta\tilde{g}_{\mu\nu}}\left(\frac{I_m}{\sqrt{-g}}\right)=0
\nonumber
\ee
Which in the Einstein frame leads us to:
\be
\nabla^\mu\nabla_\mu\phi -V_{,\phi}= \frac{1}{2}\frac{\partial \tilde{g}_{\mu\nu}}{\partial\phi_{,\alpha}}\nabla_\alpha\left(\sqrt{\frac{\tilde{g}}{g}}\tilde{T}^{\mu\nu}\right)+\frac{1}{2}\sqrt{\frac{\tilde{g}}{g}}\tilde{T}^{\mu\nu}\nabla_\alpha\left(\frac{\partial \tilde{g}_{\mu\nu}}{\partial\phi_{,\alpha}}\right)-\frac{1}{2}\frac{\partial \tilde{g}_{\mu\nu}}{\partial\phi}\left(\sqrt{\frac{\tilde{g}}{g}}\tilde{T}^{\mu\nu}\right).
\la{eqmovgen}
\ee
With no specific assumptions neither about the induced metric nor rely on any peculiar form of $\tilde{T}^{\mu\nu}$. It is not hard to generalize such a result for other specific k-essence fields. As our next step, it is convenient to discuss the role of $A(X_\phi,\phi)$ in the whole framework.
To see the relevance of this parameter, we will just assume a first order expansion in $X_\phi$ from the conformal part of the induced metric:
\be
A(X_\phi,\phi)=A^{(0)}+A^{(1)}X_\phi+O(X_\phi^2).\nonumber
\ee
In a generic scenario we must also add the terms corresponding to the disformal part.\\
From the results in \cite{Noller:2012sv} we find that [\ref{eqmovgen}] can be generally written in the Einstein frame as:
\ba
\nabla_\mu\nabla^\mu\phi&=&V_{eff,\phi}+\text{friction terms}\la{confeqmov}\\
V_{eff,\phi}&=&\frac{V_{,\phi}+A^{(0)}_{,\phi}\hat{\rho}}{1-A^{(1)}\hat{\rho}}.\la{veff}
\ea
Where $\hat{\rho}$ is the energy density that corresponds to the energy-momentum tensor conserved in the Einstein frame $\hat{T}^{\mu\nu}=A(X_\phi,\phi)\tilde{T}^{\mu\nu}$. The last result quotes a peculiar feature of this effective potential: it depends on the environment $(\hat{\rho})$. The position of the minimum is given by:
\be
V_{,\phi}(\phi_{min})+A^{(0)}_{,\phi}(\phi_{min})\hat{\rho}=0\nonumber
\ee 
And the effective mass of the field is:
\be
m^2_{eff}=V_{eff,\phi\phi}(\phi_{min})=\frac{V_{,\phi\phi}(\phi_{min})+A^{(0)}_{,\phi\phi}(\phi_{min})\hat{\rho}}{1-A^{(1)}\hat{\rho}}\nonumber
\ee
This quantity is clearly shifted (or screened) from the bare mass parameter $V_{,\phi\phi}(\phi_{c})$. We can naively give an interpretation of such a screening from the perspective of an interactive theory: the mass was expected to change because of the interaction of the field with its surroundings. It is not necessary to consider a minimum in $V(\phi)$ in order to find a lower bound in the effective potential, and consequently an environmentally dependent mass for the field. To see this, let us consider a simpler case in which $A^{(1)}(\phi)=0$, $A^{(0)}(\phi)=\alpha\phi$ and $V(\phi)=\beta/\phi^n,\:n>0$. Thus, we find an expression for the effective potential:
\be
V_{eff}(\phi)=\frac{\beta}{\phi^n}+\hat{\rho}\alpha\phi
\la{vex}
\ee
In the next figure, we illustrate the effect of the conformal factor by giving a sketch of the effective potential:

\begin{figure}[htbp]
\centering
\includegraphics[width=6cm, height=4.5cm]{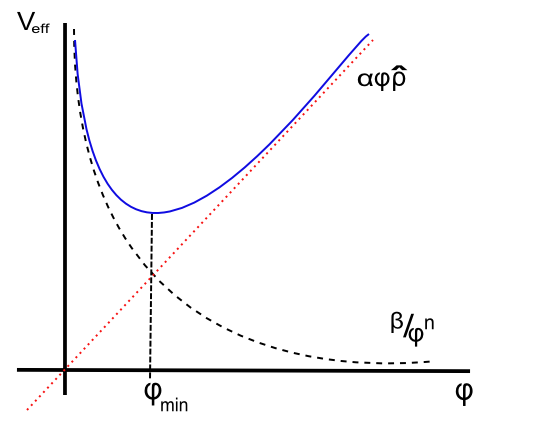}
\caption{\small{Effective potential (\textbf{solid blue}) built from a potential without minima (\textbf{dashed black}) and the conformal contribution at a fixed spacetime event (\textbf{dotted red}). Even when there is no bounded value of $\phi$ for an equilibrium configuration of $V(\phi)$, we can still find $\phi_{min}$ for the effective potential.}}
\label{fig4}
\end{figure}
\newpage
By following this procedure, we can find the value of the field which minimizes the potential and the effective mass:
\ba
\phi_{min}&=&\left(\frac{\beta n}{\alpha\hat{\rho}}\right)^{\frac{1}{n+1}}\nonumber\\
m^2_{eff}&=&\beta n(n+1)\left(\frac{\beta n}{\alpha\hat{\rho}}\right)^{-\frac{n+2}{n+1}}\nonumber
\ea
Assuming $\alpha,\beta>0$\footnote{A popular choice is $A^{(0)}(\phi)=\exp\left(\alpha\phi\right)$. With $\alpha$ arbitrarily large for a generic non-minimal coupling between the metric and the auxiliary field. In this case, the final effect is not substantially different from what we see in Figure [\ref{fig4}]}. In this simple example, we can see the explicit dependence of the two quantities with the energy density. The runaway potential $V(\phi)=\beta/\phi^n$ is the desirable for quintessence models. But in the context of this model, this is just an example of a potential with no equilibrium points.\\  
This is the essential feature of the so-called \emph{Chameleon effect}. And it is significant in our model since we do not need to impose a specific form of the potential to find an equilibrium point. Recalling [\ref{supmoddr}], we can infer that the existence of such a configuration implies that slow oscillations around the minimum in a colder universe enable subluminal propagation of matter in the Einstein frame. A more detailed insight to other possible configurations for the conformal factor can be found in \cite{Noller:2012sv}. In here, the author analyzes the static solutions of the gravitational scalar.\\ Neglecting the contribution of the disformal terms in [\ref{indmet}] at non-relativistic energy limits, and considering pressureless matter sources. The corresponding extra terms in the equations of motion are proportional to $\phi_{,\mu}\tilde{T}^{\mu\nu}$, and these vanish under the stated conditions. In addition to this, thin-shell mechanism limits the mass shifting effects for large and massive bodies ($\geq M_S$) by supressing the static solutions of the field. These facts are consistent with our expectations for the system when is close to the equilibrium. 
\subsection{An intrepid solution: A Cuscuton}\la{sectthreethree}
In the next paragraphs, we show an explicit mapping between emergent geometries and a specific type of bimetric model, by assuming a different perspective from the approach in the previous part. This is another option to describe the dynamics of the geometry inducing field. To see this, we will assume $A=1$ and $B$=constant to find an easier map between bimetric theories and the induced geometries from subsection \ref{secttwothree}. \\
Firstly, by neglecting conformal factors, we can compare [\ref{indmet}] and [\ref{indoth}], which gives us:
\be
\frac{\mathcal{L}_{,XX}}{\mathcal{L}_{,X}+2X\mathcal{L}_{,XX}}=-B.
\la{comp}
\ee
The solution of this differential equation is
\be
\mathcal{L}(X)=\frac{1}{B}\sqrt{1+2BX}-\frac{1}{B}\nonumber
\ee
If $\mathcal{L}_{,\phi}=\mathcal{L}_{,\phi X}=0$, this means that we are just dealing with the kinetic part of the Lagrangian density. Moreover, we are interested to map certain k-essence models into bimetric theories with geometries which only depend on the derivatives of the inducing field, but not on the field itself. The full functional is:
\be
\mathcal{L}(X,\phi)=\frac{1}{B}\sqrt{1+2BX}-\frac{1}{B}-V(\phi)\la{fullksessenceL}
\ee
And as we have corroborated from Appendix \ref{appone}, the perturbations of this k-essence field map can be described by Klein-Gordon action which propagates in [\ref{indmet}]  (with $A$=1). 
\newpage
It is not hard to notice in the low kinetic limit ($BX\ll 1$) we recover $\mathcal{L}(X,\phi)\approx X-V$. But it is also important to consider the opposite regime ($BX\gg 1$):
\be
\mathcal{L}(X,\phi)\approx \sqrt{\frac{2}{B}}\:\sqrt{X}+\frac{1}{\sqrt{8B^3}\:\sqrt{X}}-V(\phi)+O\left(\frac{1}{X}\right)
\la{hercusc}
\ee 
The second term (and all the following orders not considered in this expansion) is negligible compared to $\sqrt{2/B}\:\sqrt{X}$. The remaining action coincides with a Cuscuton \cite{Afshordi:2006ad}.  In Appendix \ref{apptwo}, we show the unique features of the Cuscuton solution. On the other hand, we know the determinant of [\ref{indmet}]:
\be
\tilde{g}=g(1+2BX).
\nonumber 
\ee 
And now, we suppose that the gravitational scalar is just driven by the cosmological constant, thus, rewriting the action from [\ref{fullksessenceL}], we find:
\be
\int d^4x\sqrt{-g}\mathcal{L}=\int d^4x\sqrt{-g}\left(\frac{1}{B}\sqrt{1+2BX}-\frac{1}{B}\right)=-\int d^4x\sqrt{-g}\frac{1}{B}+\int d^4x\sqrt{-\tilde{g}}\frac{1}{B}\la{kessenceact}
\ee
This is certainly a remarkable result, we only needed to add two positive constants, which are fixed by setting
\be
\frac{1}{B}=2\hat{\Lambda}.\nonumber
\ee  
This provides the correct low energy limit for the theory. For instance, we can write the full action:
\be
\int d^4x\:\sqrt{-g}\left(R(g_{\mu\nu})-2\hat{\Lambda}\right)+\int d^4x\:\sqrt{-\tilde{g}}\left(\mathcal{L}_m(\tilde{g}_{\mu\nu},\Psi_m)+2\hat{\Lambda}\right).
\la{fullcuscact}
\ee 
And as an exercise, we will write all these terms in the matter frame by considering
\be
g_{\mu\nu}=\tilde{g}_{\mu\nu}+B\phi_{,\mu}\phi_{,\nu}.\nonumber
\ee
This is nothing but the reversed version of [$\ref{indmet}$] and implies a contraction of the speed of light with respect to the broadened lightcone. By replacing it in [\ref{fullcuscact}], we get:
\be
\int d^4x\:\sqrt{-\tilde{g}}\left(\mathcal{L}_m(\tilde{g}_{\mu\nu},\Psi_m)+2\hat{\Lambda}-2\hat{\Lambda}\sqrt{1-2B\hat{X}}+\sqrt{1-2B\hat{X}}\:R(\tilde{g}_{\mu\nu})\right).\la{actdbi}
\ee 
\newpage
Where $\hat{X}=-1/2\:\tilde{g}^{\mu\nu}\phi_{,\mu}\phi_{,\nu}$. The first three terms in this expression look exactly like an action of an existing model called DBI (Dirac-Born-Infeld) inflation. It is not hard to guess that it was based on the contraction of the speed of light.
However, the gravitational term is a complete mess: it does not look like the gravitational part of that model at all. And is clearly different from any frame dependent inflationary phase. Additionally, the presence of a cosmological constant partially alleviates the perils of suggesting any specific shape of the potential. Because even when it imposes a minimum energy scale, it is still necessary to add by hand a mechanism to roll-down and oscillate. Our previous discussions (See subsection [\ref{sectthreetwo}]) suggest that non-trivial contributions from the conformal part of [\ref{indmet}] might complete the analysis.\\\\
What have we achieved so far in this chapter? The action in [\ref{fullksessenceL}] has the correct features for the realization of a cosmological model with a sensible meaning in two opposite kinetic regimes: At low kinetic contributions, the support of the Chameleon effect allows us to find a stable minima regardless of the shape of a generic k-essence potential, as long this is originally dependent of the field. This is precisely the case of the quintessence field potential in the example. And also, the extremal case of non-relativistic pressureless matter detailed in \cite{Noller:2012sv} seems to fit well in this scenario.\\ It is also important to recall [\ref{indmet}], since the without the kinetic contribution, our induced geometry is conformally related to the standard gravitational metric. This lower kinetic bound implies slow oscillations around the minimum.\\ In the opposite case, we find a Cuscuton field with infinite speed of light, free from causality issues, which couples to matter fields and enables them to propagate superluminally inside the extended lightcone. \\Furthermore, supported by the results on Appendix \ref{appone}, we find the corresponding Klein-Gordon action for perturbations of the inducing field in the matter frame. Hence, the proposal is suitable for structure formation.
\newpage
\section{Cosmological features}\la{sectfour}
\subsection{Geometric motivations}\la{sectfourone}
From the arguments exposed in Section \ref{secttwo}, we clearly notice the expansion of the matter lightcone. In addition to the results in \ref{sectthreeone}, we have enough reasons to believe that the expanded causal region solves the issues related with the Horizon and the homogeneity problems. The objective in this subsection is to consider the additional geometric effects on a timelike congruence using the induced metric of the matter sector. As an outcome of these procedures, we find that it is not necessary to violate the strong energy condition (characteristic of inflationary models) for an expanding area transversal to the congruence. In principle, this means we can reproduce the results of an inflationary model in a qualitatively different way.\\
Consider a choice of induced metric simply proportional to $g_{\alpha\beta}$ (A=1) and with a canonical kinetic contribution from the scalar $\phi$ (B=const). In that case, we can relate the connections in both metrics by
\be
\mathcal{C}^\alpha_{\beta\gamma}\equiv\tilde{\Gamma}^\alpha_{\beta\gamma}(\tilde{g}_{\alpha\beta})-\Gamma^\alpha_{\beta\gamma}(g_{\alpha\beta})=\frac{-B}{1-2BX_\phi}\tilde{\nabla}^\alpha\phi\tilde{\nabla}_\beta\tilde{\nabla}_\gamma\phi\equiv -B\:C^\alpha_{\beta\gamma}.\la{idchris}
\ee
We use the standard definition of the Riemann tensor for the induced geometry
\be
\tilde{R}^\rho_{\sigma\mu\nu}=\tilde{\Gamma}^\rho_{\nu\sigma,\mu}-\tilde{\Gamma}^\rho_{\mu\sigma,\nu}+\tilde{\Gamma}^\rho_{\mu\lambda}\tilde{\Gamma}^\lambda_{\nu\sigma}-\tilde{\Gamma}^\rho_{\nu\lambda}\tilde{\Gamma}^\lambda_{\mu\sigma},\nonumber
\ee
and after using [\ref{idchris}], the curvature tensor of the induced metric can be written as
\be
\tilde{R}^\rho_{\sigma\mu\nu}=R^\rho_{\sigma\mu\nu}+\mathcal{R}^\rho_{\sigma\mu\nu}-B\mathcal{C}^\rho_{\sigma\mu\nu}.\la{rmod}
\ee
Where
\ba
\mathcal{R}^\rho_{\sigma\mu\nu}&=&\mathcal{C}^\rho_{\nu\sigma,\mu}-\mathcal{C}^\rho_{\mu\sigma,\nu}+\mathcal{C}^\rho_{\mu\lambda}\mathcal{C}^\lambda_{\nu\sigma}-\mathcal{C}^\rho_{\nu\lambda}\mathcal{C}^\lambda_{\mu\sigma}\nonumber\\
\mathcal{C}^\rho_{\sigma\mu\nu}&=&\Gamma^\rho_{\lambda[\mu}C^\lambda_{\nu]\sigma}+C^\rho_{\lambda[\mu}\Gamma^\lambda_{\nu]\sigma}\la{coup}
\ea
Considering the action in [\ref{fullcuscact}], we find the field equations
\be
R_{\mu\nu}-\frac{1}{2}g_{\mu\nu}R+\hat{\Lambda}g_{\mu\nu}=8\pi G\sqrt{\frac{\tilde{g}}{g}}\tilde{T}_{\mu\nu},\la{feqcusc}
\ee
where the cosmological constant in the matter sector has been absorbed by $\tilde{T}_{\mu\nu}$. Using the last expression, we write the Ricci tensor in terms of the energy-momentum tensor as a conserved current in the matter frame:
\be
\small{R_{\mu\nu}=8\pi G\sqrt{\frac{\tilde{g}}{g}}\left[\left(\tilde{T}_{\mu\nu}-\frac{\tilde{T}}{2}\tilde{g}_{\mu\nu}\right)-\frac{B}{2}\tilde{T}\tilde{\nabla}_\mu\phi\tilde{\nabla}_\nu\phi-\frac{B}{1+2B\tilde{X}_\phi}\left(\tilde{T}_{\alpha\beta}\tilde{\nabla}^\alpha\phi\tilde{\nabla}^\beta\phi\right)g_{\mu\nu}\right]+2\hat{\Lambda}g_{\mu\nu}}.\la{ricci}
\ee
Again $\tilde{\nabla}^\alpha$ is the covariant derivative with respect to the new metric. It is relevant to recall that we are evaluating the strong energy condition. As such, this is only meaningful in the matter geometry because of the definition in [\ref{trinv}].
The evolution of a timelike congruence of curves is given by the Raychaudhuri equation:
\be
\dot{\theta}+\frac{1}{3}\theta^2=-\tilde{R}_{\mu\nu}\tilde{u}^\mu\tilde{u}^\nu-2\left(\sigma^2-\omega^2\right)\la{raych}
\ee
Replacing [\ref{ricci}] and the contracted [\ref{rmod}], we have
\ba
\dot{\theta}+\frac{1}{3}\theta^2&=&-8\pi G\sqrt{\frac{\tilde{g}}{g}}\left(\tilde{T}_{\mu\nu}-\frac{\tilde{T}}{2}\tilde{g}_{\mu\nu}\right)\tilde{u}^\mu\tilde{u}^\nu+4\pi G\sqrt{\frac{\tilde{g}}{g}}B\tilde{T}\left(\tilde{\nabla}_\mu\phi\tilde{u}^\mu\right)^2\nonumber\\
&+&\frac{8\pi G B}{1+2B\tilde{X}_\phi}\sqrt{\frac{\tilde{g}}{g}}\left(\tilde{T}_{\alpha\beta}\tilde{\nabla}^\alpha\phi\tilde{\nabla}^\beta\phi\right)g_{\mu\nu}\tilde{u}^\mu\tilde{u}^\nu-\mathcal{R}_{\mu\nu}\tilde{u}^\mu\tilde{u}^\nu\nonumber\\
&+& B\mathcal{C}^\rho_{\mu\rho\nu}\tilde{u}^\mu\tilde{u}^\nu- 2\hat{\Lambda}g_{\mu\nu}\tilde{u}^\mu\tilde{u}^\nu-2\left(\sigma^2-\omega^2\right).\la{raychfin}
\ea
And in order to achieve an irrotational expanding congruence $\left(\dot{\theta}+1/3\:\theta^2>0\right)$, the latter expression leads us to the following condition
\ba
\left(\tilde{T}_{\mu\nu}-\frac{\tilde{T}}{2}\tilde{g}_{\mu\nu}\right)\tilde{u}^\mu\tilde{u}^\nu&<& \frac{B}{2}\tilde{T}\left(\phi_{,\mu}\tilde{u}^\mu\right)^2+\frac{B}{1+2B\tilde{X}_\phi}\left(\tilde{T}_{\alpha\beta}\tilde{\nabla}^\alpha\phi\tilde{\nabla}^\beta\phi\right)g_{\mu\nu}\tilde{u}^\mu\tilde{u}^\nu\la{cond}\\&-&\frac{1}{8\pi G}\sqrt{\frac{g}{\tilde{g}}}\:\mathcal{R}_{\mu\nu}\tilde{u}^\mu\tilde{u}^\nu+\frac{B}{8\pi G}\sqrt{\frac{g}{\tilde{g}}}\:\mathcal{C}^\rho_{\mu\rho\nu}\tilde{u}^\mu\tilde{u}^\nu-\frac{\hat{\Lambda}}{4\pi G}\sqrt{\frac{g}{\tilde{g}}}g_{\mu\nu}\tilde{u}^\mu\tilde{u}^\nu\nonumber
\ea 
The limit in which [\ref{fullksessenceL}] becomes a KG action implies that the cosmological constant term in the RHS is balanced with the ground energy level absorbed by the terms in the left. Additionally, $\mathcal{R}_{\mu\nu}\approx 0$ for spatially homogeneous solutions of $\phi$, and $g_{\mu\nu}\tilde{u}^\mu\tilde{u}^\nu>0$ for superluminal expansion. Thus, in contrast with the focusing theorem, we are not obliged to break the strong energy condition to pursue an expanding behavior. The last result agrees with most of the literature about the subject, specially with the results in \cite{Clayton:1998hv} and \cite{Kouretsis:2012ys}. This is a key feature to understand the results coming in the next chapter of this manuscript.
\subsection{Modified FLRW cosmology}
In this subsection, we reproduce the results in \cite{Clayton:1999zs} and find them coherent with the results in the last section. To be more concrete, spatial homogeneity is assumed for the field. The results in [\ref{sectthreetwo}] are useful to remember that in principle we do not need pick a specific shape if we add a nontrivial contribution of the conformal factor $A$ in [\ref{indmet}]. And because of that, to simplify our discussions, we will limit the action of this conformal term just to assume an effective potential with a defined equilibrium position. Leaving $A=1$ for all other purposes.\\
Under these conditions, the modified line element suggested is
\be
ds^2=-\left(1+B\dot{\phi}^2\right)dt^2+a^2(t)\left[\frac{dr^2}{1-kr^2}+r^2d\Omega^2\right],
\la{flrwmod}
\ee
where the gravitational metric is given by the standard FLRW coordinates. Furthermore, this expression for the metric is useful to find the scale $\beta$ introduced in [\ref{modlorentz}], for the modified Lorentz transformations in Section \ref{sectthree}
\be
\beta(t)=\frac{1}{\sqrt{1+B\dot{\phi}^2}}.\nonumber
\ee
And from here, we notice that this relation for $\beta$ corresponds to a spatially homogeneous widening of the lightcone in the induced geometry.\\  
From [\ref{flrwmod}], we find
\be
\sqrt{-\tilde{g}}=\left(1+B\dot{\phi}^2\right)\sqrt{-g}\la{detmetfrw}.
\ee
As a simple model for matter, we consider a perfect fluid
\be
\tilde{T}^{\mu\nu}=\left(\rho+P\right)u^\mu u^\nu-p\tilde{g}^{\mu\nu}
\la{fluidmat}
\ee
with a vector field normalized by $\tilde{g}_{\mu\nu}u^\mu u^\nu=-1$ such that
\be
u^0=1/\sqrt{1+B\dot{\phi}^2}.\nonumber
\ee
We can also write the conservation equation from [\ref{defT}]
\be
\dot{\rho}+3\frac{\dot{a}}{a}\left(\rho+P\right)=0.
\la{cons}
\ee
Knowing this, the field equation in [\ref{eqmovgen}] reduces to
\be
\left(1-\frac{16\pi GB}{(1+B\dot{\phi}^2)^{3/2}}\rho\right)\ddot{\phi}+3\frac{\dot{a}}{a}\dot{\phi}\left(1+\frac{16\pi GB}{\sqrt{1+B\dot{\phi}^2}}P\right)+V_{,\phi}(\phi)=0.\nonumber
\ee 
From [\ref{genact}] in the limit case $\mathcal{L}_\phi=X-V$, we find the Friedmann equations for the system:
\ba
\left(\frac{\dot{a}}{a}\right)^2+\frac{k}{a^2}&=&\frac{\Lambda}{3}+\frac{1}{6}\left[\frac{\dot{\phi}^2}{2}+V(\phi)\right]+\frac{8\pi G}{3}\frac{\rho}{\sqrt{1+B\dot{\phi}^2}}\la{fried1}\\
\left(\frac{\dot{a}}{a}\right)^2+2\frac{\ddot{a}}{a}+\frac{k}{a^2}&=&\Lambda-\frac{1}{2}\left[\frac{\dot{\phi}^2}{2}-V(\phi)\right]-8\pi G\sqrt{1+B\dot{\phi}^2}P.\la{fried2}
\ea
Subtracting [\ref{fried1}] from [\ref{fried2}], we find
\be
\frac{\ddot{a}}{a}=\frac{\Lambda}{3}+\frac{V(\phi)}{3}-\frac{\dot{\phi}^2}{6}-\frac{8\pi G}{3}\left(\frac{\rho}{\sqrt{1+B\dot{\phi}^2}}+3\sqrt{1+B\dot{\phi}^2}P\right).\nonumber
\ee
Hence, we must write the last expression in the comoving time coordinate defined in the matter frame
\be
\tau\equiv\int\left(1+B\dot{\phi}^2\right)^{1/2}dt,\nonumber
\ee
using $K=1+B\dot{\phi}^2$ as an auxiliary variable, and with assignments of ``energy density'' and ``pressure'' for the scalar field similar to [\ref{defsc}], we find
\be
\frac{{a}''}{H^2a}=\frac{K'}{2HK}-\frac{1}{2}(1+K\Omega_k)-\frac{4\pi G}{H^2}\left(p_\phi+\sqrt{K}P-\frac{\Lambda}{8\pi G}\right) \la{acc}.
\ee 
Where $a'\equiv da/d\tau$, $H\equiv a'/a$ and $\Omega_k\equiv k^2/H^2$. The first Friedmann equation in [\ref{fried1}] can be rewritten as
\be
1+K\Omega_k=K\Omega_\Lambda+\Omega_\phi+K^{3/2}\Omega_M,\la{friedmod}
\ee
defining $8\pi G\rho/3H^2=\Omega_M$, $8\pi G\rho_\phi/3H^2=\Omega_\phi$ and  $\Lambda/3H^2=\Omega_\Lambda$.\\
The expression in [\ref{acc}] is fully compatible with the results of the previous part: if we assume that the interactions of the scalar field are dominant over the matter interactions (which is the ``moral'' of [\ref{cond}]) we achieve a regime of expansion. With zero cosmological constant, we get
\be
\frac{K'}{2HK}>\frac{4\pi G}{H^2}\left(p_\phi+3\rho_\phi+3\rho+\sqrt{K}P\right),\la{condinf}
\ee
regardless of the content of the matter sector. However, from these expressions we can also notice that if the matter sector is dominated by species that violate strong energy condition, the inflationary effect is enhanced.\\
Our next task is to describe the Horizon problem in the context of this model. According to [\ref{flrwmod}], the formula for the comoving size of the particle horizon at the time of last scattering $t^*$ is
\be
d_H(t^*)=\int_0^{t^*}\frac{\sqrt{1+B\dot{\phi}^2}}{a(t)}dt.
\la{phor}
\ee
We can use a similar expression to estimate the distance to the last scattering surface
\be
d_{LS}=\int_{t^*}^{t_0}\frac{\sqrt{1+B\dot{\phi}^2}}{a(t)}dt,\la{lastsc}
\ee 
where $t_0$ denotes the present epoch. To solve the horizon problem, we need the particle horizon at $t^*$ to be larger than the distance between two opposite points in the sky, which means
\be
d_H(t^*)\geq 2d_{LS}.\la{horpro}
\ee 
An equivalent statement can be phrased by evaluating [\ref{phor}] at these days
\be
d_H(t_0)=d_H(t^*)+d_{LS}\leq \frac{3}{2}d_H(t^*).\nonumber
\ee 
This can be easily accomplished by just by considering $1+B\dot{\phi}^2\gg 1$ while $0\leq t\leq t^*$, this was deliberately intended in the model building developed in Section \ref{secttwo}.\\
To get a more precise value of $1+B\dot{\phi}^2$, we learned from \cite{Moffat:1998mp} that we can use the first Friedmann equation [\ref{fried1}] written in terms of relative abundances at the superluminary radiation era:
\be
|\Omega(10^{-43} \text{sec})-1|\sim O\left(\left(1+B\dot{\phi}^2\right)\times 10^{-60}\right).\nonumber
\ee
This is not significantly different from the usual fine-tuning problem we face in standard cosmology, for $1+B\dot{\phi}^2\sim 10^{58}$, we find
\be
|\Omega(10^{-43} \text{sec})-1|\sim O\left(1\right),\nonumber
\ee
which implies much less fine-tuning than the standard FLRW model.\\
Since we look to fulfill the condition in [\ref{cond}] and [\ref{condinf}], the roughest of all approximations allows equate the right hand side of [\ref{acc}] to a positive constant $(\gamma^2/H^2)$. Thus, we have:
\be
a(\tau)\sim \: \exp\left(\gamma\tau\right).\la{flat}
\ee 
And for $\gamma>0$ this implies a possible solution to the flatness problem. However, the transition to low energy scales returning to standard Lorentz symmetry is not a trivial problem faced by this model. We will discuss some of these issues in the next paragraphs.
\newpage
\subsection{A graceful exit?}
So far in this project, we have described a model which tries to be consistent at different energy scales. And also, we find logical connections with the process of structure formation from perturbations of the scalar field, as depicted in Appendix \ref{appone}. Nevertheless, there is an additional fact that has been oblivious in the arguments used so far: if we consider the gravitational frame in this model, a potential present configuration of the Universe still implies very small deviations from the exact dispersion relations in Special Relativity. In addition to this, the results in \cite{Clayton:1999zs} seem to suggest that the theory might be extended well inside the electroweak symmetry breaking regime. In the revision process of this project, we encountered several objections for this statement.\\ In \cite{Collins:2004bp}, \cite{Polchinski:2011za} and others, it has been stated that such deviations can lead us into unphysical effects, described by the percolation of Lorentz violating terms into low energy scales. We can find a specific example in \cite{Collins:2004bp}. In this paper, the authors calculate the self-energy diagram of a fermion in a Yukawa theory. Modifying the free fermion propagator by a smooth factor $f(|\mathbf{k}|/\Lambda)$, with $f(\infty)=0$ and $f(0)=1$. Finding a non-negligible Lorentz violation parameter:
\be
\xi=\frac{g^2}{6\pi^2}\left[1+2\int_0^\infty xf'(x)^2 dx\right].\nonumber
\ee 
This correction corresponds to a first order modification to the speed of light \footnote{A comment: we recall [\ref{supmoddr}], where we have found quadratic deviations with repect to the inducing field. Which are certainly weaker than the stated in the reference.}. There have been many proposals trying to solve this pathology, including the insertion of higher dimensional operators and custodial symmetries (SUSY has been proposed as a candidate in \cite{GrootNibbelink:2004za} and \cite{Bolokhov:2005cj}). And certainly, the residual oscillations of the inducing field force us to quote that this model is not exempt from future analysis and similar corrections. We did not use any other specific (local or rigid) symmetry than the translational invariance in the matter frame. Nevertheless, when the action is rewritten in this frame (in analogy to [\ref{actdbi}]), these issues seem to be diluted in the matter sector $\left(\tilde{g}_{\mu\nu}=\eta_{\mu\nu}\right)$, transferring the extra degrees of freedom to the gravitational side. But in this project, we do not intend to provide a full discussion on these affairs.\\  
\section{Discussions}
\par In this project, our approach to Bimetric Models of Gravity was utilitarian in a cosmological context. Nonetheless, the relevance of these models has been exploited in the context of Massive Gravity \cite{Baccetti:2012bk}. And also, there are interesting attempts to think on black holes in the context of these models in \cite{Deffayet:2011rh}.\\
\par We discussed the effects of considering modified dispersion relations which are still quadratic. The mechanism of induction described in Section \ref{secttwo} made use of the essentials of Finsler geometry. However, it is very clear from the example of Section \ref{secttwoone} that the knowledge of these techniques extends these results to consider aquadratic terms in the dispersion relations. Which motivate other interesting scenarios such as \cite{Magueijo:2002xx} ($f(p_0=E)=1+p_0/M$, $M$ plays the role of the Planck scale). It is worthwhile to mention the contribution made by the authors in \cite{Kouretsis:2012ys}, where we can find a cogent discussion about the peculiar features of Finsler geometries. Many of these are specially relevant to cosmological models.\\
\par As we mentioned before, we just focused our attention in quadratic modifications. For instance, a constrained application of the induction method used for a generic Finsler geometry had as an outcome the disformal Riemannian metric found in [\ref{indmet}]\footnote{Recently, a different approach followed by  Kothawala in \cite{Kothawala:2013maa} introduces the notion of a disformal metric with results similar to [\ref{indmet}].}. The constraint appeared when we picked the invariants in [\ref{scinv}] instead of other contracted quantities. Naively, we can observe the analogy between the induction method used in the example of \ref{secttwoone} and the emergent geometry in Subsection \ref{secttwothree}. Behind this analogy, it is not hard to notice that the equations of motion of a generic k-essence field do not look like a typical Klein-Gordon equation: these are the ``field version'' of a modified dispersion relation. The quest for the inducted geometry is an attempt to look for a frame in which the equations of motion of a k-essence field can be written as a KG equation.\\ Once we found the induced metric, we looked for the right set of conditions to obtain a broadened lightcone in both perspectives.\newpage
We give special attention to the fact that $B>0$ since this is a crucial difference with other existing models (such DBI Inflation or MOND).\\
\par The geometric description of the matter sector for this model is followed by a profile of the dynamics of the inducing fields in Section \ref{sectthree}. It is important to mention that the expansion in the matter lightcone is meaningless if the matter modes are not able to propagate beyond the gravitational lightcone. And because of that, the effect of the scalar is illustrated with an example in \ref{sectthreeone} in the gravitational frame, with the only purpose to show that ordinary matter can travel superluminally inside the expanded causal region. Nevertheless, the appearance of instabilities is expected in this frame, casting doubts about frame dependence on many of the statements made so far.\\
The fact that measurements and experiments are conducted in the matter frame (using rods and clocks made of matter) is an argument commonly used to justify these results. If the reader is still not convinced by this idea, we can find an interesting insight to this issue in \cite{Bruneton:2006gf}.\\ In this article, the author suggests that many of this issues are related to the imposition of a preferred chronology of events in either of this frames. Proposing the introduction of a global (mixed) chronology to avoid a biased choice.\\
\par Going further in this manuscript, we developed the case $\mathcal{L}_\phi=X-V$ in order to introduce the effects of the conformal factor in the model. This case is also important since it is a natural lower kinetic limit for a generic k-essence Lagrangian density. The \emph{Chameleon effect} provides an alternative to find at least a minimum in the (environmentally dependent) effective potential, regardless of any preliminary choice of a potential in the k-essence model. In addition to this, we present the ideas in \cite{Magueijo:2008sx} in order to give a simple map between the two methods of induction described in Section \ref{secttwo}. In [\ref{actdbi}], we see the inverse of such a map describes a model different than any other inflationary proposal. The expression in [\ref{fullksessenceL}] considers a Klein-Gordon action in the lower kinetic regime $(BX_\phi\ll 1)$ and a Cuscuton action in the opposite case $(BX_\phi\gg 1)$. The solutions of the field in the last regime holds a vast spectrum of conceptual details, superficially reviewed in Appendix \ref{apptwo}.\\
\newpage
\par According to \cite{Magueijo:2008sx}, $\phi$ is described by the KG equations in the matter frame when this is driven just by the cosmological constant. We believe that the potential is not naturally excluded: The presence of a conformal factor in the induced metric plays a relevant role by transforming the ordinary field potential into an expression with well-defined minima. However, the result remains valid for oscillations around the minimum, or a nearly flat potential. In either of these cases, the action follows straightforwardly from the results in Appendix \ref{appone}. Guilelessly, we believe that the instability of the k-essence field in this scenario might not be seen as harmful.\\
\par In the last section of this paper, we developed a minimally modified FLRW cosmology. Firstly, we explored the conditions for an expanding congruence in the matter frame. In this reference, we have evaluated the strong energy condition, finding that the violation of this is not mandatory for an accelerated phase. In contrast with the results in \cite{Bassett:2000wj}, where the breaking of this condition is stated as mandatory. In this article, the conserved currents and charges were defined in different frames than the assumed throughout this project. However, we can notice from [\ref{raychfin}] and [\ref{acc}] that the imposition of such violations enhances an inflationary phase. We have found enough evidence in [\ref{cond}] to reaffirm the subdominance of matter interactions during the expansion. The dominance comes from the interactions of the inducing field with gravity, in our calculations this can be found in
\be
C_{grav-\phi}=\frac{B}{8\pi G}\sqrt{\frac{g}{\tilde{g}}}\:\mathcal{C}^\rho_{\mu\rho\nu}\tilde{u}^\mu\tilde{u}^\nu\nonumber
\ee
at the right hand side of [\ref{cond}]. It is reasonable to expect a pronounced contribution from this term in a regime of Quantum Gravity. Moreover, the condition found in [\ref{condinf}] remarks the dominance of the kinetic terms (time derivatives of the inducing field) for the expansion. This is certainly opposite to the slow-roll behavior expected from a standard inflaton. 
\par In the last subsection of this chapter, we briefly introduced some residual effects taken into account generated by the breaking of Lorentz symmetry at low energy scales in the gravitational frame. A survey of this violations and its implications is developed in \cite{Liberati:2013xla}. Where we can also find observational limits and tests which can be applied for this model.
\begin{appendices}
\section{Action for perturbations}\la{appone}
Here we discuss the properties of the motion equation in [\ref{defothmet}]. Our treatment follows the ideas in \cite{Babichev:2007dw} and \cite{Magueijo:2008sx} and aims to map the perturbations of a k-essence theory into a Klein-Gordon picture in the matter frame. To do so, let us rewrite the equation as follows:
\be
G^{\mu\nu}\nabla_\mu\nabla_\nu\phi-\rho_{,\phi}=J
\la{eqapp}
\ee
Where the covariant derivatives are written in terms of the gravitational metric and $J$ comes from the coupling with other degrees of freedom in the full action. Now we must consider the variation of this expression caused by the splitting of the field solutions $\phi(x,t)=\phi_0+\delta\phi$
\be
G^{\mu\nu}\nabla_\mu\nabla_\nu\delta\phi-\rho_{,\phi\phi}\delta\phi-\rho_{,\phi X}\nabla_\mu\phi_0\nabla^\mu\delta\phi+\left(\frac{\partial G^{\mu\nu}}{\partial\phi}\delta\phi+\frac{\partial G^{\mu\nu}}{\partial\nabla_\alpha\phi}\nabla_\alpha\delta\phi\right)\nabla_\mu\nabla_\nu\phi_0=\delta J
\la{vareqapp}
\ee
This variation can be rearranged as
\be
G^{\mu\nu}\nabla_\mu\nabla_\nu\delta\phi+V^\mu\nabla_\mu\delta\phi-\tilde{M}^2\delta\phi=\delta J,
\la{varcomp}
\ee
where
\ba
V^\mu&\equiv&\frac{\partial G^{\alpha\beta}}{\partial\nabla_\mu\phi}\nabla_\alpha\nabla_\beta\phi_0-\rho_{,\phi X}\nabla^\mu\phi_0\nonumber\\
\tilde{M}^2&\equiv&\frac{\partial G^{\mu\nu}}{\partial\phi}\nabla_\mu\nabla_\nu\phi_0-\rho_{,\phi\phi}\la{appdef}
\ea
We want to achieve a Klein-Gordon equation in a metric conformally related with $G_{\mu\nu}$
\be
\tilde{G}^{\mu\nu}D_\mu D_\nu\delta\phi-M^2\delta\phi=\delta\tilde{J},
\la{modkg}
\ee
this implies a redefined covariant derivative 
\be
D_\mu A_\nu=\nabla_\mu A_\nu-C_{\mu\nu}^\alpha A_\alpha\la{covder}
\ee
We express [\ref{modkg}] in more familiar terms to compare it with [\ref{varcomp}]:
\ba
\Omega G^{\mu\nu}\nabla_\mu\nabla_\nu\delta\phi+\Omega V^\mu\nabla_\mu\delta\phi-\Omega\tilde{M}^2\delta\phi&=&\Omega\delta J\nonumber\\
\tilde{G}^{\mu\nu}\nabla_\mu \nabla_\nu\delta\phi-\tilde{G}^{\mu\nu}C_{\mu\nu}^\alpha\nabla_\alpha\delta\phi-M^2\delta\phi&=&\delta\tilde{J}\la{comp}
\ea
Leading us to identify $\Omega G^{\mu\nu}=\tilde{G}^{\mu\nu}$, $\Omega\tilde{M}^2=M^2$, $\Omega\delta J=\delta\tilde{J}$ and $\Omega V^\alpha=-\tilde{G}^{\mu\nu}C_{\nu\mu}^\alpha$. The last identity is a defining property of the connection:
\be
\tilde{G}^{\mu\nu}C_{\mu\nu}^\alpha=-\Omega\left(\frac{\partial G^{\mu\nu}}{\partial\nabla_\alpha\phi}\nabla_\mu\nabla_\nu\phi_0-\rho_{,\phi X}\nabla^\alpha\phi_0\right)\la{conn}
\ee
Using the analogous of an identity well-known in GR
\be
\tilde{G}^{\mu\nu}C_{\mu\nu}^\alpha=-\frac{1}{\sqrt{-\tilde{G}}}\nabla_\beta(\sqrt{-\tilde{G}}\tilde{G}^{\alpha\beta})
\la{trgr}
\ee 
we wish to find the precise value of $\Omega$ such that it satisfies the identification made on [\ref{comp}]. And to do so, we must calculate the determinant of $G^{\mu\nu}$. By the definition of this induced metric [\ref{defothmet}], we find:
\be
\text{det}\left(\mathcal{L}_{,X}g^{\mu\nu}-\mathcal{L}_{,XX}\phi^{,\mu}\phi^{,\nu}\right)=\text{det}\left(\mathcal{L}_{,X}g^{\mu\nu}\right)\text{det}\left(\delta^\mu_\alpha-\mathcal{L}_{,X}^{-1}\mathcal{L}_{,XX}\phi^{,\mu}\phi^{,\nu}g_{\alpha\nu}\right)\nonumber
\ee
Using $\text{det}\left(e^A\right)=e^{trA}$ we have:
\be
\text{det}\exp\left[\ln\left(\delta^\mu_\alpha-\mathcal{L}_{,X}^{-1}\mathcal{L}_{,XX}\phi^{,\mu}\phi^{,\nu}g_{\alpha\nu}\right)\right]=\exp\left[\text{tr}\left(\ln\left(\delta^\mu_\alpha-\mathcal{L}_{,X}^{-1}\mathcal{L}_{,XX}\phi^{,\mu}\phi^{,\nu}g_{\alpha\nu}\right)\right)\right]\la{idapp}
\ee
Expanding the logarithm in taylor series:
\ba
\text{tr}\left[\ln\left(1-\mathcal{L}_{,X}^{-1}\mathcal{L}_{,XX}\left(\phi^{,\mu}\phi^{,\nu}\mathbf{g}^{-1}\right)\right)\right]&=&\sum\limits_{k}\frac{(-1)^{k+1}}{k}\left[\mathcal{L}_{,X}^{-1}\mathcal{L}_{,XX}\right]^k \text{tr}\left(-\phi^{,\mu}\phi^{,\nu}\mathbf{g}^{-1}\right)^k\nonumber\\&=&\sum\limits_{k}\frac{(-1)^{k+1}}{k}\left[2X\mathcal{L}_{,X}^{-1}\mathcal{L}_{,XX}\right]^k\nonumber\\&=&\ln\left(1+2X\mathcal{L}_{,X}^{-1}\mathcal{L}_{,XX}\right)\nonumber
\ea
Replacing in [\ref{idapp}] we finally get:
\be
\text{det}\left(G^{\mu\nu}\right)=\mathcal{L}_{,X}^{4}c_s^{-2}\text{det}(g^{\mu\nu})\nonumber
\ee
And the inverse is simply $\text{det}\left(G^{\mu\nu}\right)^{-1}$. With respect to $\tilde{G}$:
\be
\text{det}\left(\tilde{G}^{\mu\nu}\right)=\Omega^{4}\mathcal{L}_{,X}^{4}c_s^{-2}\text{det}(g^{\mu\nu})
\ee
If we define the auxiliary function $F=\sqrt{-\tilde{G}}\Omega/\sqrt{-g}$ and also consider [\ref{trgr}], we can rewrite [\ref{conn}] as follows:
\be
\nabla_\lambda\left(FG^{\alpha\lambda}\right)=F\left(\frac{\partial G^{\mu\nu}}{\partial\nabla_\alpha\phi}\nabla_\mu\nabla_\nu\phi_0-\rho_{,\phi X}\nabla^\alpha\phi_0\right)\la{finexp}
\ee
By applying the chain rule, this is equivalent to:
\be
G^{\alpha\lambda}\nabla_\lambda F=F\left(\left(\frac{\partial G^{\mu\nu}}{\partial\nabla_\alpha\phi_0}-\frac{\partial G^{\mu\alpha}}{\partial\nabla_\nu\phi_0}\right)\nabla_\mu\nabla_\nu\phi_0-\left(\frac{\partial G^{\lambda\alpha}}{\partial\phi}+\rho_{,\phi X}g^{\alpha\lambda}\right)\nabla_\lambda\phi_0\right)\nonumber
\ee
A careful derivation of all the terms in the right hand side of this equation allows us to have as a final expression
\be
G^{\alpha\lambda}\nabla_\lambda F=0.\nonumber
\ee
The auxiliary function $F$ is a constant, which can be chosen to be 1. And by using its definition we find the final value $\Omega$
\be
\Omega=\frac{c_s}{\mathcal{L}_{,X}^{2}}\la{omega}
\ee 
This factor completely determines the geometry in which the dynamics of perturbations can be described by the Klein-Gordon action:
\ba
I_{\delta\phi}&=&\int d^4x\sqrt{-\tilde{G}}\:\left(-\frac{1}{2}\tilde{G}^{\mu\nu}D_\mu\delta\phi D_\nu\delta\phi-\frac{M^2}{2}\delta\phi^2\right)\nonumber\\
\tilde{G}^{\mu\nu}&=&\frac{c_s}{\mathcal{L}_{,X}^{2}}\left(\mathcal{L}_{,X}g^{\mu\nu}-\mathcal{L}_{,XX}\phi^{,\mu}\phi^{,\nu}\right)\nonumber
\ea 
Notice that the \emph{new} covariant derivatives in [\ref{covder}] appear in the action. The conformal factor does not alter causal structure of the induced spacetime. In here, the procedure is slightly more general than the one followed by \cite{Magueijo:2008sx}. We can get exactly the same results by considering $\mathcal{L}_{,\phi}=\mathcal{L}_{,\phi X}=0$, leading us directly to a massless Klein-Gordon system.  This result is extremely important to understand how the perturbations of a k-essence field can be described using the induced geometry of the matter frame. In principle it is not compulsory to have a Klein-Gordon picture of the perturbations in the matter frame, but is undeniable that this is certainly convenient to study structure formation. And therefore, present tangible evidence to compare with another proposed models.
\newpage
\section{\emph{Standard} features of a Cuscuton field}
\begin{figure}[htbp]
\centering
\includegraphics[width=5.5cm, height=5cm]{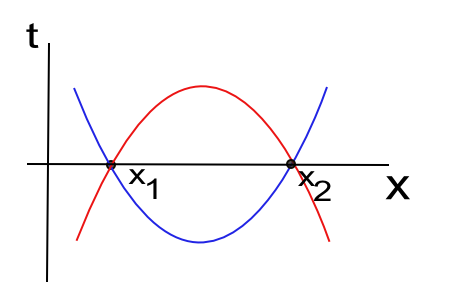}
\caption{\small{In this picture, we illustrate the only two possible configurations (in red and blue) for $\phi_0$ under the imposed boundary conditions. Depending on the position of the centering coordinates ($t_0,x_0$).}}
\label{fig5}
\end{figure}

\la{apptwo}
Motivated by \cite{Afshordi:2006ad}, we discuss some peculiar aspects of the Cuscuton field provided by the high energy limit of [\ref{hercusc}]. Firstly, we present the equations of motion for the field. Then, we use a particular case to find a geometric interpretation of the solutions, including the discrete nature of them. We also give consistent arguments to show the absence of internal dynamics in the Cuscuton and the collapse of the phase space. Concluding with a brief discussion about the infinite group velocity of the perturbations.\\
Just keeping the terms of [\ref{hercusc}] relevant to our analysis, we find 
\be
\mathcal{L}(X,\phi)= \sqrt{\frac{2}{B}}\:\sqrt{X}-V(\phi).
\la{extcusc}
\ee
The equations of motion are simple:
\be
\nabla_\mu\left(\frac{\phi^{,\mu}}{\sqrt{-\phi^{,\mu}\phi_{,\mu}}}\right)+\sqrt{B}V_{,\phi}(\phi)=0
\la{laeqmovcusc}
\ee
We can define a normalized vector
\be
n^{\mu}\equiv\frac{\phi^{,\mu}}{\sqrt{-\phi^{,\mu}\phi_{,\mu}}}
\ee
and this can be interpreted as the normal vector of a constant field surface. This interpretation simplifies our understanding of [\ref{laeqmovcusc}], which is equivalent to
\be
K^{\mu}_\mu=-\sqrt{B}V_{,\phi}(\phi)
\ee
where $K^{\mu}_\nu$ is the extrinsic curvature tensor. And it means that hypersurfaces of constant $\phi$ have constant mean curvature (CMC). Many features of the explicit solutions for the field can be studied in 1+1 Minkowski spacetime, in this case the field equation reduces to
\be
\left(t-t_0\right)^2-\left(x-x_0\right)^2=K^{-2}=\frac{1}{BV_{,\phi}(\phi)^2}
\la{simpcusc}
\ee
A euclidean version of this equation can be illustrated as a sphere (or a soap bubble). With this result, we argue that we can only obtain a discrete set of possible solutions.
\newpage
Suppose we know $\phi_0(t=0,x)$ which is the initial condition for the field, fixing the curvature for the initial hyperbola (See Figure [\ref{fig5}]). And we also consider the boundary conditions for the hyperbola at $x=x_1$ and $x=x_2$: every solution $\phi(t,x)$ must pass through both points at $t=0$, this just allows two possible hyperbolae to be considered as an initial set up for the system.
However, this is very far from a formal proof of this conjecture. If we consider all space dimensions, the initial configuration might not be a sphere. In that case we will need a more general procedure.\\
Another interesting property of this field is that it has no local dynamics. Therefore, a Cuscuton fluid carries zero entropy. To see this, we transform the phase space from the Lagrangian ($D\phi\wedge D\dot{\phi}$) to the Hamiltonian measure ($D\phi\wedge Dp$).\\
Firstly, we calculate the conjugate momentum:
\be
p(\dot{\phi})=\frac{1}{\sqrt{B}}\int d^3x\:\sqrt{-g}\:\frac{\phi^{,0}}{\sqrt{-\phi^{,\mu}\phi_{,\mu}}}\la{cuscmom}
\ee
The Jacobian for this transformation is
\be
\text{det}\left(\frac{\partial \phi(x)p(x')}{\partial \phi(y) \dot{\phi}(y')}\right)=\text{det}\left(\begin{array}{cc}
\delta^{3}(\mathbf{x}-\mathbf{y}) & 0 \\ 
\alpha & \frac{\left(\nabla\phi\right)^2}{\left(\sqrt{-B^{\frac{1}{3}}\phi^{,\mu}\phi_{,\mu}}\right)^3}\delta^{3}(\mathbf{y}-\mathbf{y'}) \\ 
\end{array}\right).
\ee
It is always possible to rotate $\phi_{,\mu}$ in such a way that the spatial gradient vanishes $(\nabla\phi)^2=0$ and cancels the Jacobian.
\newpage
The same argument works in the matter frame being specially careful with spacetime dependence of the rotation parameters. Hence, the basics of statistical mechanics point out that this theory has a collapsed phase space (zero volume, no accessible states) and no internal dynamics. Notice that the discrete nature of the solutions are best viewed as points in the phase space, implying a measure equal to zero.\\ Even under these conditions, there are no restrictions on the interactions of this field with ordinary matter for exactly the same reason. If we remember the calculations made in subsection [\ref{sectthreeone}], the potential adds nonvanishing  off-diagonal terms, and the determinant of the 4$\times$4 Jacobian matrix will not be zero. The argument of Lorentz symmetry will not hold in this case: a transformation to the rest frame of the Cuscuton will not have the same effect in a generic matter field since we proposed a change of scale for the transformations in the matter frame.\\ The bottom line of these arguments is that Cuscuton field might not be used to send information, but it can be seen as a \emph{vehicle} for other fields to propagate superluminally.\\\\ Besides, from [\ref{laeqmovcusc}] we find the corresponding equations of motion for perturbations:
\be
\delta\phi^{,\mu}_{,\mu}+n^\mu n^\nu\delta\phi_{,\mu\nu}+\sqrt{B}\:V''(\phi_0)\left(\sqrt{-\phi_0^{,\mu}\phi_{0,\mu}}\right)\delta\phi\nonumber
\ee
Analyzing the static solutions in the Fourier domain, we find:
\be
\left(\omega^2-k_{||}^2\right)\delta\phi_k+\sqrt{B}\:V''(\phi_0)\left(\sqrt{-\phi_0^{,\mu}\phi_{0,\mu}}\right)\delta\phi_k=0.
\la{parall}
\ee
Where $k_{||}$ is the momentum component that remains parallel to the CMC surface. Confirming the idea of a propagation restricted to a 3-D surface, this is another evidence of the lack of internal dynamics.\\\\
And finally, considering [\ref{extcusc}], a direct application of [\ref{cs}] allows us to find a diverging speed of sound:
\be
\left(1+2X\frac{\mathcal{L}_{,XX}}{\mathcal{L}_{,X}}\right)=0\longrightarrow c_s^2\rightarrow\infty
\la{divcs}
\ee
This is not seen as a problem since we already noticed that the trajectories followed by a Cuscuton carry no information. Naively, these results suggest many motivations to call the Cuscuton an \emph{aether field}. 
\end{appendices}
\section*{Acknowledgements}
I am indebted to my supervisor Carlo Contaldi, for the patience and the guidance necessary to complete this dissertation. And also I would like to thank Amel Durakovic, Ghazal Geshnizjani, David Romero, Sarah Shandera and Teofilo Vargas for comments on earlier drafts and many fruitful discussions about the subject.

\end{document}